\documentclass[10pt,journal,compsoc]{IEEEtran}

\makeatletter
\def\mdseries@tt{m}
\makeatother

\newif\ifblind
\blindfalse

\usepackage[utf8]{inputenc}
\usepackage{graphicx}
\usepackage{xcolor}
\usepackage{xspace}
\definecolor{kellygreen}{RGB}{0, 161, 98}
\definecolor{kellyred}{RGB}{191, 13, 2}
\usepackage[colorlinks=true,citecolor=kellygreen,linkcolor=kellyred]{hyperref}
\usepackage{booktabs}
\usepackage{tikz}
\usepackage{wrapfig}

\usepackage[frozencache=true]{minted}
\fvset{linenos, fontsize=\scriptsize, frame=lines, numbersep=1pt, xleftmargin=.75em}
\AtBeginEnvironment{minted}{%
  }

\usepackage[per-mode=symbol]{siunitx}
\DeclareSIUnit \bit {bit}
\DeclareSIUnit \bits {bits}
\DeclareSIUnit \byte {B}
\DeclareSIUnit \bytes {Bytes}
\DeclareSIUnit \cycle {cycle}
\DeclareSIUnit \cycles {cycles}
\DeclareSIUnit \hz {Hz}
\DeclareSIUnit \op {Op}
\DeclareSIUnit \ops {Ops}
\DeclareSIUnit \operand {operand}
\DeclareSIUnit \operands {operands}
\DeclareSIUnit \transfer {T}
\DeclareSIUnit \cell {cell}

\title{Python FPGA Programming\\with Data-Centric Multi-Level Design}

\newcommand{\fblas}{\textsc{fBLAS}\xspace}

\author{%
\ifblind
\IEEEauthorblockN{Anonymous authors.}
\else
\author{\IEEEauthorblockN{Johannes de Fine Licht$^*$, Tiziano De Matteis$^*$, Tal Ben-Nun$^*$, Andreas Kuster$^*$\\Oliver Rausch$^*$, Manuel Burger$^*$, Carl-Johannes Johnsen$^{\dagger*}$, Torsten Hoefler$^*$}\\[0.5em]
\IEEEauthorblockA{$^*$Department of Computer Science, ETH Zurich, $^\dagger$Department of Computer Science, University of Copenhagen
}}
\fi
}

\IEEEtitleabstractindextext{%
\begin{abstract}
Although high-level synthesis (HLS) tools have significantly improved programmer productivity over hardware description languages, developing for FPGAs remains tedious and error prone.
Programmers must learn and implement a large set of vendor-specific syntax, patterns, and tricks to optimize (or even successfully compile) their applications, while dealing with ever-changing toolflows from the FPGA vendors.
We propose a new way to develop, optimize, and compile FPGA programs.
The Data-Centric parallel programming (DaCe) framework allows applications to be defined by their dataflow and control flow through the Stateful DataFlow multiGraph (SDFG) representation, capturing the abstract program characteristics, and exposing a plethora of optimization opportunities.
In this work, we show how extending SDFGs with multi-level Library Nodes incorporates both domain-specific and platform-specific optimizations into the design flow, enabling knowledge transfer across application domains and FPGA vendors.
We present the HLS-based FPGA code generation backend of DaCe, and show how SDFGs are code generated for either FPGA vendor, emitting efficient HLS code that is structured and annotated to implement the desired architecture.
\end{abstract}}

\begin{document}

\maketitle

\section{Introduction}
\label{sec:introduction}


The widespread adoption of HLS tools have greatly improved programmer productivity when targeting FPGAs by allowing kernels to be developed in C++ or OpenCL~\cite{hls_for_fpgas}. Still, achieving efficient architectures remains challenging in practice, as the optimization space of hardware design is larger than for software; requiring new syntax, new ways to structure programs, and new transformations~\cite{hls_transformations}. In this work, we propose employing the \emph{Data-Centric parallel programming} (DaCe) framework and its \emph{Stateful DataFlow multiGraph (SDFG)}~\cite{dace} intermediate representation as an alternative way to develop FPGA programs. The data-centric representation enables explicit management of data location and movement, which remains the biggest performance factor in computing today~\cite{data_locality}.

SDFGs allow representing programs by their dataflow and control flow independent of the chosen FPGA backend, enable compatibility across FPGA vendors through code generation, and are amenable to optimizing transformations performed directly on the graph. SDFGs have been proven effective for load/store workloads in various domains, ranging from linear algebra kernels and graph algorithms~\cite{dace} to numerical weather prediction~\cite{stencilflow} and supercomputer-scale quantum transport simulations~\cite{omen}. When FPGA SDFGs are manually authored~\cite{dace}, their performance is on-par with state-of-the-art implementations and libraries.

Modern compiler techniques, such as polyhedral optimization, allow sophisticated automatic optimization of low-level IR by detecting and transforming loop constructs~\cite{pluto}, but is restricted to the space of transformations that can be proven to be ``safe''. Rather than relying on fully automated optimization to exploit all available opportunities, DaCe exposes powerful performance analysis capabilities and optimization tools that enable knowledgeable performance engineers to perform \emph{guided} optimization of programs. Transformations are done via graph rewriting on the SDFG representation, expressed in terms of the general dataflow and control flow of the program, thus facilitating knowledge exchange between programs and domains.

Domain-specific languages can enable additional optimizations by restricting the input domain, allowing additional assumptions to be made on the program's behavior, but typically offer limited exchange of knowledge and engineering effort with other domains.
In this work, we show how the ``Library Node'' extension to SDFGs, first prototyped by StencilFlow~\cite{stencilflow}, enable a multi-level design methodology that exposes the best of both worlds within the same framework, enabling the application of both domain-specific and general purpose optimizations to DaCe programs.

\begin{figure}[t]
    \centering
    \includegraphics[width=.99\columnwidth]{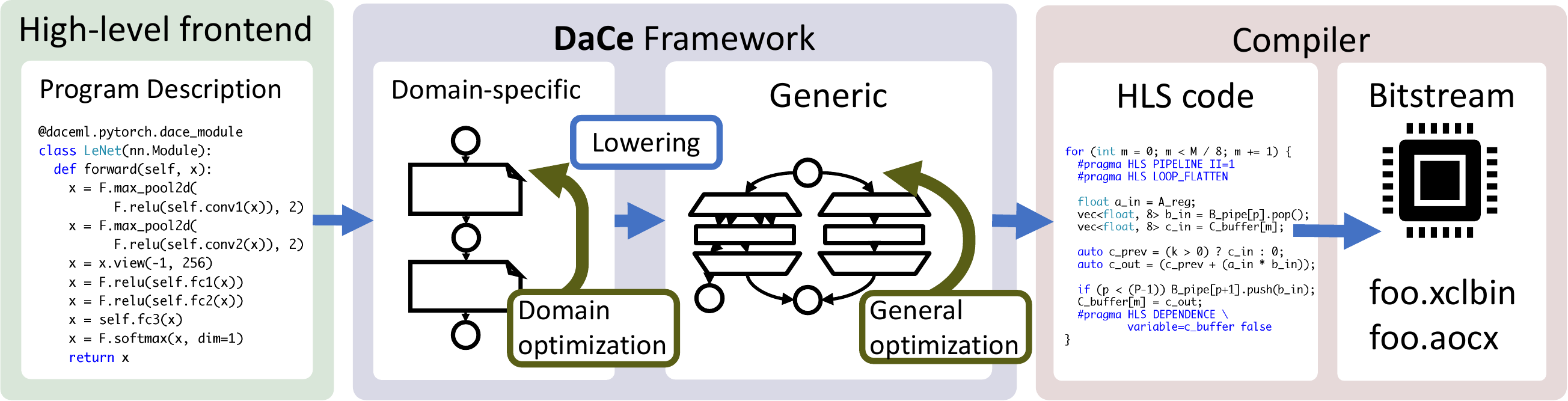}
    \caption{Proposed development workflow using the DaCe framework.}
    \label{fig:workflow}
\end{figure}

\noindent Throughout this work, we describe how the proposed methodology enables productively developing fast FPGA programs by utilizing a multi-level approach:
\begin{itemize}
    \item SDFGs natively expose key aspects of hardware, such as pipelining, streaming, and systolic arrays.
    \item DaCe's FPGA backend targets both Xilinx and Intel FPGAs with structured, annotated HLS code, enabling reuse between vendors.
    \item DSL frontends, such as BLAS, ONNX, or StencilFlow~\cite{stencilflow}, allow expressing programs at a high abstraction level, comprehensible to non-FPGA experts.
    \item High-level domain-specific transformations allow applying domain knowledge to optimize programs.
    \item General mid-level transformations allow porting from CPU to FPGA, and perform key optimizations, such as replacing memory accesses with streaming.
    \item Low-level specialization allow employing vendor-specific tweaks, targeting non-universally supported features such as shift registers or accumulators.
\end{itemize}

\section{Representation and Code Generation}
\label{sec:dace_fpga_backend}

\begin{figure}[t]
    \centering
    \includegraphics[width=.99\columnwidth]{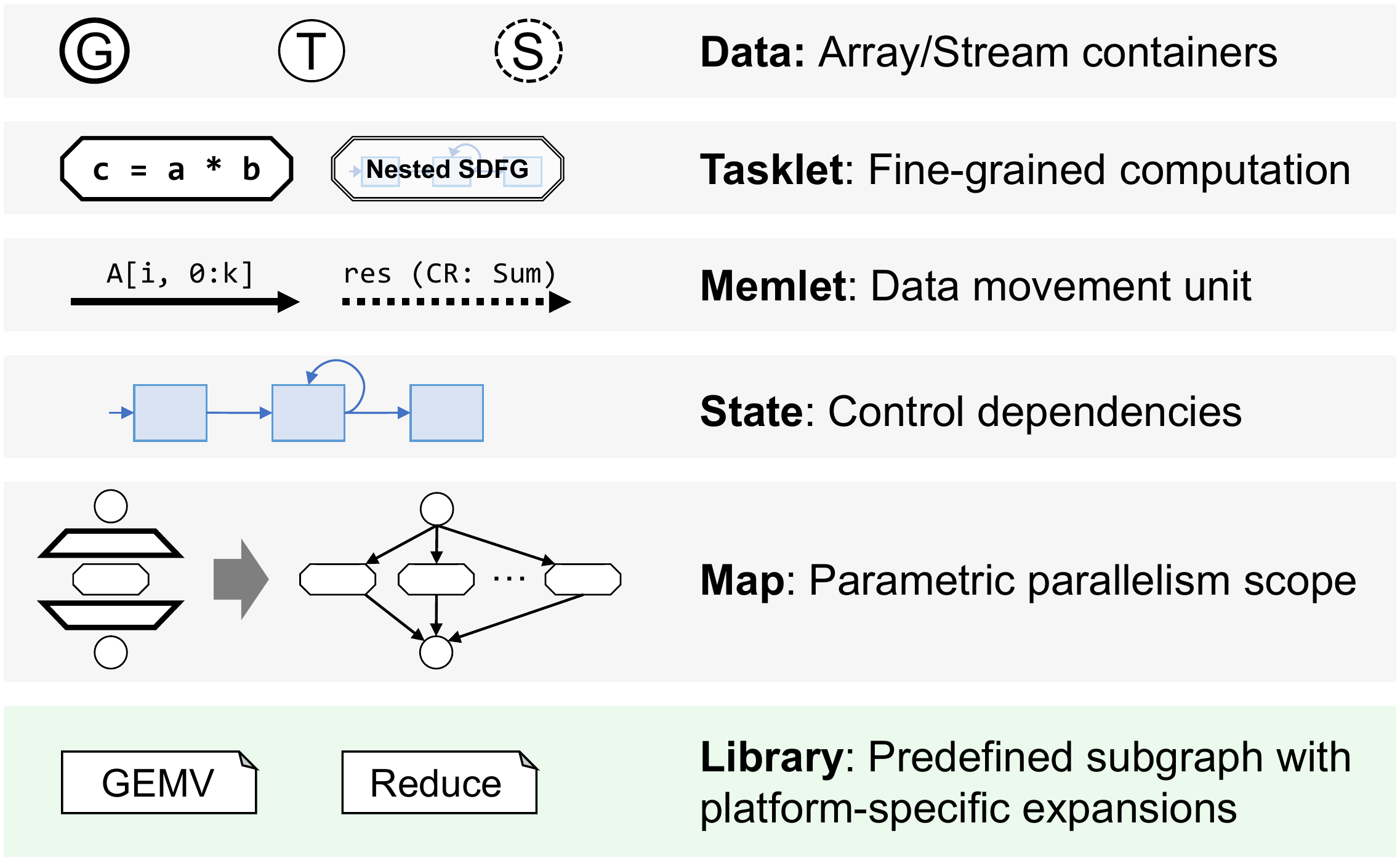}
    \caption{Glossary of nodes and edges in the SDFG representation. This work extends SDFGs with Library Nodes, highlighted in green.}
    \label{fig:sdfgs}
\end{figure}

In the following, we provide an overview of how SDFGs represent FPGA programs, and how key concepts are translated into fast HLS code for the two major vendors: Xilinx, through the Vivado~HLS~\cite{autopilot} C++-based compiler; and Intel, through the Intel OpenCL SDK for FPGA~\cite{altera_opencl}.
The components of SDFGs that we will use are summarized in Figure~\ref{fig:sdfgs}, and we give a brief primer here.

Access nodes represent \textbf{data containers} in dataflow states of the SDFG (drawn as ovals with solid (arrays) or dashed (streams) borders), where they are accessed using \textbf{memlets}, annotated on dataflow edges. Computations are performed in small tasks called \textbf{tasklets} (drawn as hexagons), which \emph{can only access memory explicitly passed to them via dataflow edges}, capturing \emph{all} data movement in the program. \textbf{Maps} implement explicit parallelism in the graph, representing parametric replication of the content between map \emph{entry} nodes (opening the scope) and map \emph{exit} nodes (closing the scope), drawn as trapezoids. Coarse-grained control flow is represented by control flow graphs, where nodes are \textbf{states} that only contain pure dataflow. Control flow and dataflow graphs can be arbitrarily nested in each other using \textbf{nested SDFGs} to represent arbitrary program semantics, while still exposing as much analyzable data movement as possible. Operational semantics can be found in the initial paper~\cite{dace}.

\begin{figure}[b]
    \centering
    \begin{tikzpicture}
        \node[anchor=south west,inner sep=0] at (0, 0) {\includegraphics[width=.9\columnwidth]{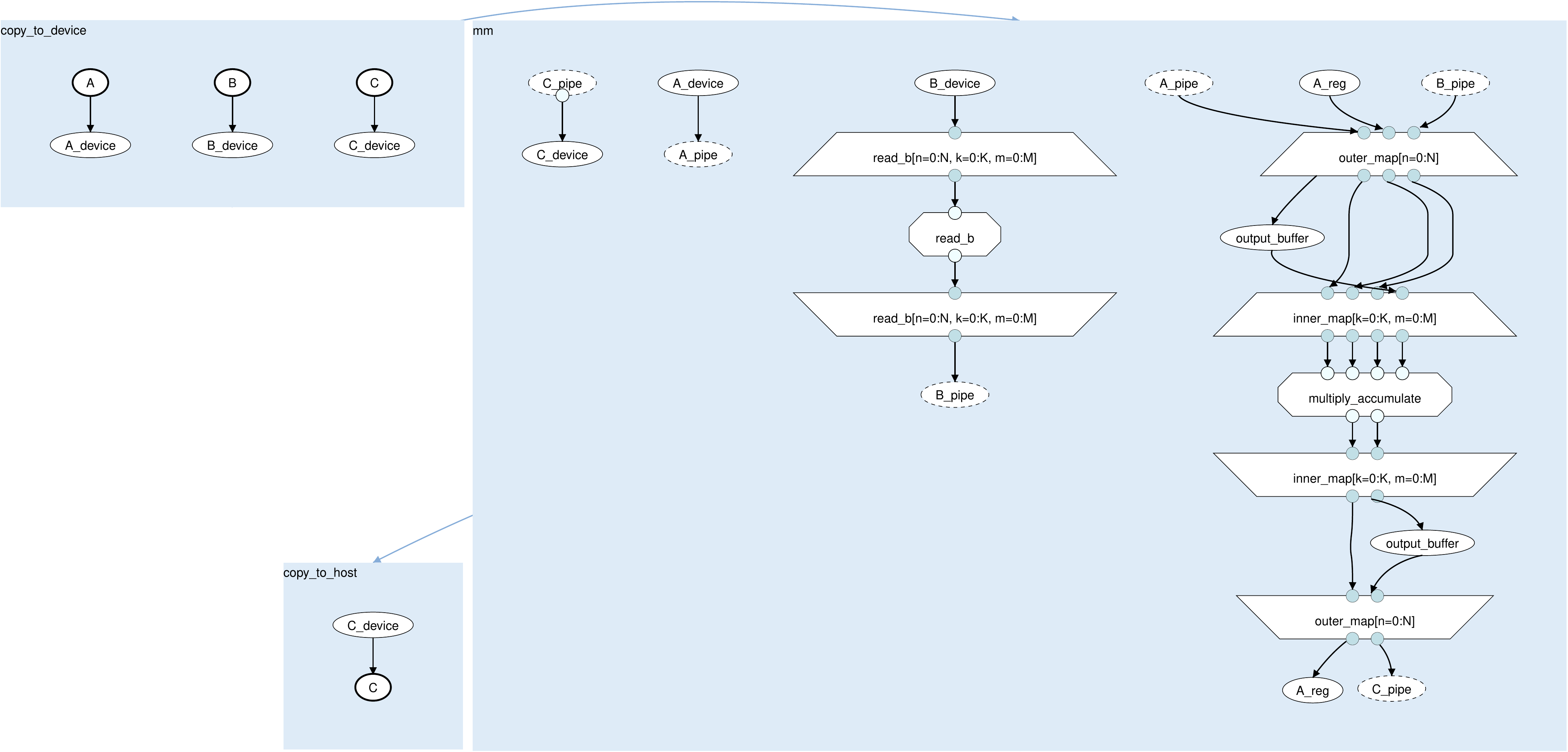}};
        \draw [blue, thick, rounded corners, dashed] (2.60, 2.8) rectangle (3.15, 3.6);
        \draw [red, thick, rounded corners, dashed] (3.30, 2.8) rectangle (3.85, 3.6);
        \draw [black, thick, rounded corners, dashed] (4, 1.65) rectangle (5.75, 3.6);
        \draw [purple, thick, rounded corners, dashed] (5.85, 0.15) rectangle (7.85, 3.6);
    \end{tikzpicture}
    \caption{Kernel state with four processing elements (right), with pre- and post-states (left) copying memory between host and device.}
    \label{fig:fpga_kernel}
\end{figure}
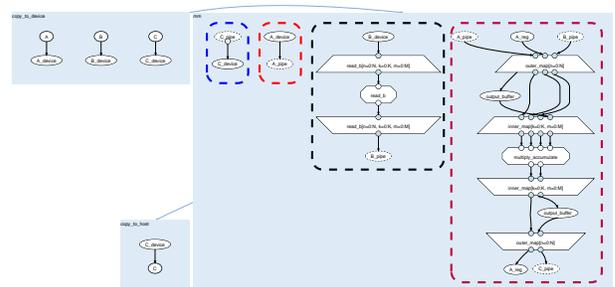

This work uses \textbf{Library Nodes}, a special version of the tasklet representing \emph{abstract behavior}, which is ``expanded''/``specialized''/``lowered'' (we will use these functionally equivalent terms interchangeably) into parametric \emph{subgraphs} that implement the abstract behavior \cite{stencilflow}. An example SDFG targeting FPGA is shown in Figure~\ref{fig:fpga_kernel}, which we will cover in detail in the following sections.

\subsection{Code Generating SDFGs}
\label{sec:code_generation}

SDFGs are constructed, manipulated, and compiled using the DaCe framework. DaCe follows the guiding principle that as many optimization opportunities as possible should be kept part of the representation --- where they can be manipulated by the performance engineer --- rather than happening as ``magic'' during code generation. Nevertheless, emitting functional and efficient code from SDFGs poses a significant design and engineering challenge, with numerous kinks and subtleties arising from moving from software to the hardware domain. The code generator must translate the final representation into structured HLS code that is easily digestible by the compiler, faithfully follows the functional semantics of the SDFG, and that successfully achieves all parallelism implied by the representation.

The FPGA backend of DaCe is modularized into a generic part, which orchestrates the traversal of the SDFG, and two lower level specialized backends for Xilinx and Intel, which are responsible for emitting vendor specific code for Vivado~HLS (C++) and the Intel OpenCL compiler, respectively. The generic backend contains the most sophistication in terms of interpreting the representation and delegating code generation tasks, whereas the two specialized components are primarily concerned with vendor-specific semantics (e.g., how processing elements and memory interfaces are expressed) and syntax (e.g., vector data types and stream objects). In particular, the highly restricted syntax supported by OpenCL requires more verbose syntax to be emitted than for backends that support C++.
DaCe also supports tasklets with embedded RTL code: in this case, both HLS and RTL kernels are generated, and then combined together in the final bitstream \cite{double_pump}.

\begin{figure*}[hbt]
    \begin{minipage}[b]{.365\textwidth}
        \begin{minted}{C++}
void mm(float *A, float *B, float *C,
        int n) {
  // ...interface pragmas omitted...
  #pragma HLS DATAFLOW
  DATAFLOW_INIT();
  dace::FIFO<float, 1, 4> A_pipe[P+1];
  dace::FIFO<float, 1, 1> B_pipe[P+1];
  dace::FIFO<float, 1, 1> C_pipe[P+1];
  DATAFLOW_FUNCTION(read_A, A, A_pipe, n);
  DATAFLOW_FUNCTION(read_B, B, B_pipe, n);
  for (size_t p = 0; p < P; p += 1) {
    #pragma HLS UNROLL
    DATAFLOW_FUNCTION(compute, p, A_pipe,
                      B_pipe, C_pipe, n);
  }
  DATAFLOW_FUNCTION(write_C, C, C_pipe, n);
  DATAFLOW_FINALIZE();
}
        \end{minted}
        \caption{Processing elements are function calls in a top-level function in the Vivado~HLS paradigm.}
        \label{fig:processing_element_xilinx}
    \end{minipage}\hfill%
    \begin{minipage}[b]{.33\textwidth}
        \begin{minted}{C++}
hlslib::ocl::Kernel kernels[] = {
    program.MakeKernel("read_A", A, n),
    program.MakeKernel("read_B", B, n),
    program.MakeKernel("compute", n),
    program.MakeKernel("compute_1", n),
    program.MakeKernel("compute_2", n),
    program.MakeKernel("compute_3", n),
    program.MakeKernel("write_C",
                       C, n)};
std::vector<cl::Event> events = {
    kernels[0].ExecuteTaskFork(),
    kernels[1].ExecuteTaskFork(),
    kernels[2].ExecuteTaskFork(),
    kernels[3].ExecuteTaskFork(),
    kernels[4].ExecuteTaskFork(),
    kernels[5].ExecuteTaskFork(),
    kernels[6].ExecuteTaskFork()};
cl::Event::waitForEvents(events);
        \end{minted}
        \caption{Processing elements are launched from host code in the Intel OpenCL paradigm.}
        \label{fig:processing_element_intel}
    \end{minipage}\hfill%
    \begin{minipage}[b]{.295\textwidth}
        \centering
        \begin{tikzpicture}
            \node[anchor=south west,inner sep=0] at (0, 0) {\includegraphics[width=.875\columnwidth]{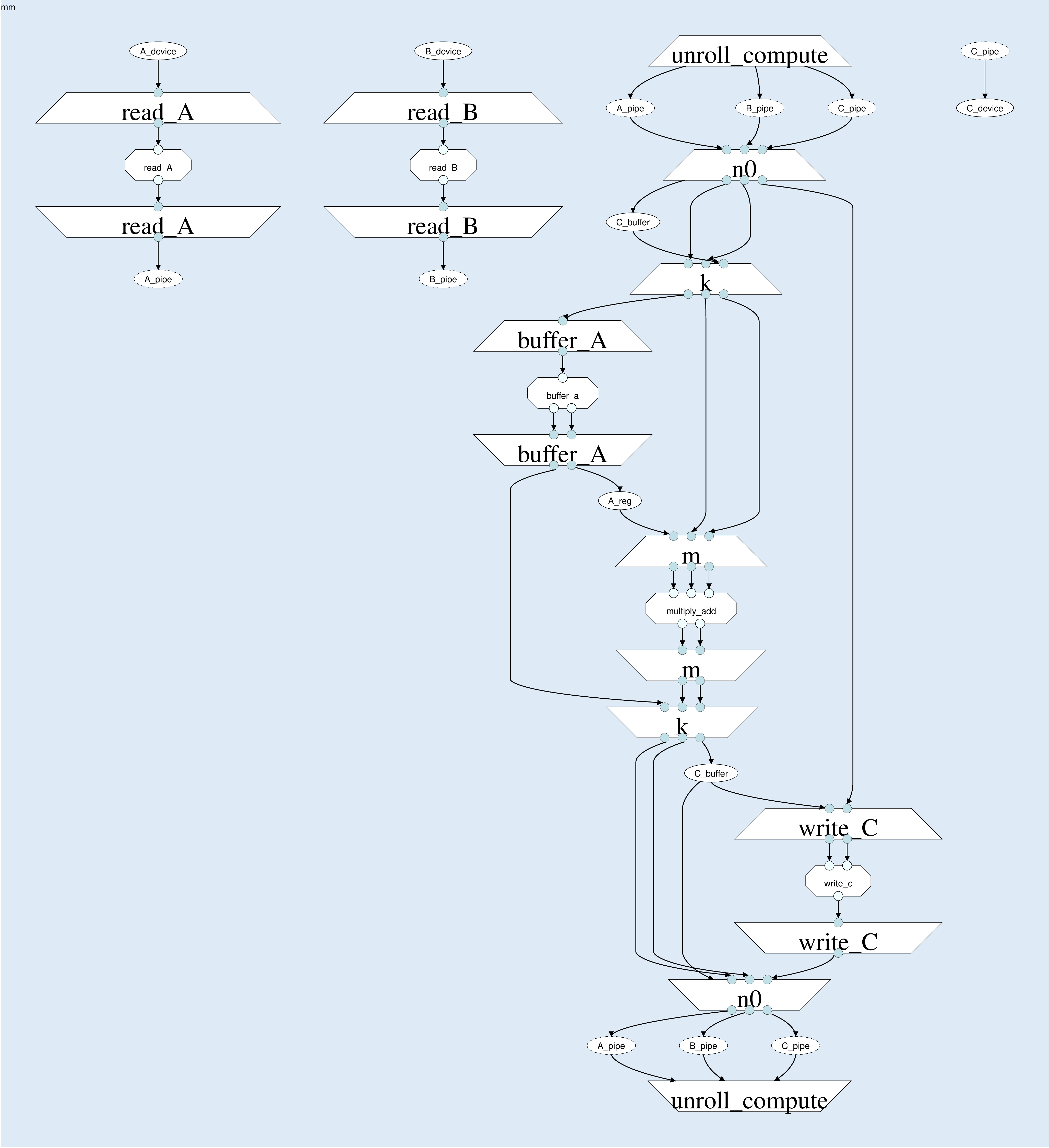}};
            \draw [red, ultra thick, rounded corners, dashed] (2.75, 4.75) rectangle (4, 5.05);
            \draw [red, ultra thick, rounded corners, dashed] (2.75, 0.1) rectangle (4, 0.4);
            \fill [blue, opacity=0.075] (2.1, 0.4) rectangle (4.25, 4);
            \fill [blue, opacity=0.075] (2.65, 4) rectangle (4.25, 4.75);
        \end{tikzpicture}
        \caption{Multiple processing elements, including a parametrically sized systolic array.}
        \label{fig:systolic_array}
    \end{minipage}
\end{figure*}

\subsection{Parallelism, Pipelining, and Unrolling with Maps}
\label{sec:pipelining}

\textbf{Representation.} Parallel sections in SDFGs are expressed via the \emph{map} construct, appearing as a pair of entry/exit nodes opening/closing scopes (one map is present in the black box, and two in the purple box in Figure~\ref{fig:fpga_kernel}). In software, these scopes can be used to target multi-core and SIMD parallelism for both CPUs and GPUs. In hardware, we distinguish between two ways of exploiting the parallelism implied by maps: \emph{pipelined} maps, where iterations are executed in sequence, but exploit pipeline parallelism in the mapped computation; and \emph{unrolled} maps, which represent parametrically replicated hardware, such as systolic arrays (see Section~\ref{sec:systolic_arrays}) or SIMD-style vectorization. The purple box in Figure~\ref{fig:fpga_kernel} contains an inner map, which will be generated as a pipelined inner loop, and an outer map over tiles, orchestrating the buffering behavior.

\textbf{Code generation.} In an SDFG with any number of nested map scopes, code generation follows the philosophy that anything that is not explicitly unrolled should be pipelined. The Intel OpenCL compiler does this automatically, but for the Xilinx backend, the graph is traversed from outermost to innermost nesting to detect the innermost map \emph{that is not unrolled}, where a pipeline pragma will be injected inside the generated loop. Furthermore, loop coalescing pragmas are automatically injected whenever loops generated from maps are perfectly nested, and when necessary, pragmas to ignore dependencies (see Section~\ref{sec:memory_hierarchy}). Maps designated as being unrolled will annotate the generated loops with the vendor-specific unroll directive.

\subsection{Representing FPGA Kernels}

\textbf{Representation.} The pure dataflow representation of SDFG states is a natural fit for mapping to streaming dataflow kernels on FPGA. When traversing the SDFG, the framework detects states \emph{that only access memory situated on the FPGA}, designating these as FPGA kernels. Although kernels are always mapped from pure dataflow states, coarse-grained control flow is still achievable by embedding \emph{nested SDFGs} in the dataflow state.
Moving data between the host and device is represented as memory copies in the representation. Access nodes are annotated with a data \emph{location}. When connected by direct data-to-data edges in a dataflow state, this will result in the appropriate copy operation depending on both source and destination. Streaming transfers can be natively represented using stream access nodes, but due to the OpenCL abstraction adopted by both the Xilinx and Intel toolflows, the backend currently only supports bulk transfers. Host/device streaming will be introduced once either backend exposes sufficient support to end-users (e.g., using the QDMA~\cite{qdma} subsystem for Xilinx FPGAs).

\textbf{Code generation.} When the code generation traversal encounters an FPGA kernel according to the aforementioned storage and execution predicate, the dataflow section is dispatched to the FPGA backend. Before continuing the traversal to generate the hardware itself, the kernel ``boundary'' is generated by inferring the necessary arguments that must be passed to the resulting OpenCL kernel launch(es). Interaction with the OpenCL API is wrapped in the interface provided by the hlslib~\cite{hlslib} C++ library, as shown in Figure~\ref{fig:processing_element_intel}.

\subsection{Processing Elements}
\label{sec:processing_elements}
\label{sec:pes}

\textbf{Representation.} The notion of partitioning the functionality of a kernel into multiple independently-scheduled modules, commonly referred to as \emph{processing elements} (PEs), is central to designing large FPGA architectures. Native support for this concept is thus a core consideration in the SDFG representation. At the same time, this should not introduce new FPGA-specific concepts to the representation.

SDFG states imply pure dataflow by representing data movement and data dependencies (e.g., everything contained in a blue rectangle Figure~\ref{fig:fpga_kernel} or Figure~\ref{fig:systolic_array}), the latter of which must be respected by the code generating backend. When a dataflow graph contains more than one \emph{weakly connected component} (i.e., at least two subgraphs $G_0$ and $G_1$ with no dataflow edge $(u, v)$ connecting any node $u{\in}G_0$ with any node $v{\in}G_1$), the backend has the liberty to \emph{schedule each weakly connected component in parallel}. For software backends, this can enable launching multiple concurrent GPU kernels, or running different concurrent tasks on multiple CPU threads. When appearing within an FPGA kernel, these are also scheduled as independent ``tasks'', exposing the concept of processing elements to the programmer. In the example in Figure~\ref{fig:fpga_kernel}, each of the four connected components represent an independent processing element scheduled in parallel: the components in the red and the black box are memory reader/prefetcher modules, which read from arrays (solid borders) in off-chip memory into data streams (dashed borders). The red box is a simple copy of the full array dimensions, implemented by a single dataflow edge, where the black box repeats multiple reads of the array, using a map to generate the desired access pattern. The blue box inversely writes from a stream back to memory.

\textbf{Code generation.} In the Vivado~HLS toolflow, processing elements are expressed by annotating a scope in the C++ code with the \texttt{DATAFLOW} pragma, resulting in every loop and function call in the scope to be scheduled as a distinct processing element. This requires a top-level ``entry'' function that contains the processing elements, is annotated with additional pragmas that designate the hardware interfaces used by the kernel to interact with the FPGA shell, and instantiates the on-chip streams (i.e., FIFOs) that facilitate inter-PE communication, shown for an example in Figure~\ref{fig:processing_element_xilinx}. The Xilinx backend uses the simulation extensions from hlslib~\cite{hlslib} (providing the \texttt{HLSLIB\_DATAFLOW\_FUNCTION} macro wrappers and the thread-safe and bounded \texttt{hlslib::Stream} class) to achieve \emph{actually} concurrent simulation of parallel processing elements, including support for feedback/back-edges in the dataflow. The Intel OpenCL flow takes a different approach: rather than being contained in a top-level function, every processing element must be expressed as a separate OpenCL kernel in the top-level scope, where they are connected using global \texttt{channel} objects. Launching each processing element is thus done from the host code, shown in Figure~\ref{fig:processing_element_intel}. These two methods of expressing kernels thus affect both the host code (which kernels are generated and launched) and the kernel code (one vs. multiple top-level kernels, global channel objects vs. local stream objects). If a generated OpenCL kernel has \emph{no} arguments, it will be generated as an ``\texttt{autorun}'' kernel, which is always active and will run whenever data is available on the connected channels, and thus does not need to be invoked from the host code.

\subsection{Channels/Streams/FIFOs}
\label{sec:streams}
\label{sec:fifos}
\label{sec:channels}

\textbf{Representation.} Streams are a native data container construct in the SDFG representation, representing first-in, first-out queues, which can be used to communicate between subgraphs in dataflow sections. In CPU and GPU codes, these are employed as single- or multi-producer queues: for example, a breadth-first search kernel can produce tasks to a queue that is consumed by multiple workers, dynamically distributing work. Stream semantics are the same in FPGA kernels, but with additional constraints due to the underlying hardware implementation that they imply: streams cannot be unbounded, and must be single-producer, single-consumer. Streams facilitate communication between processing elements, while simultaneously acting as \emph{synchronization} primitives between kernels through the producer-consumer relationship. Even though the components in Figure~\ref{fig:fpga_kernel} do not have dataflow edges between them, they synchronize by pushing/popping the same stream data container.

\begin{figure}[t]
    \centering
    \begin{minipage}{.49\columnwidth}
        \includegraphics[width=.99\columnwidth]{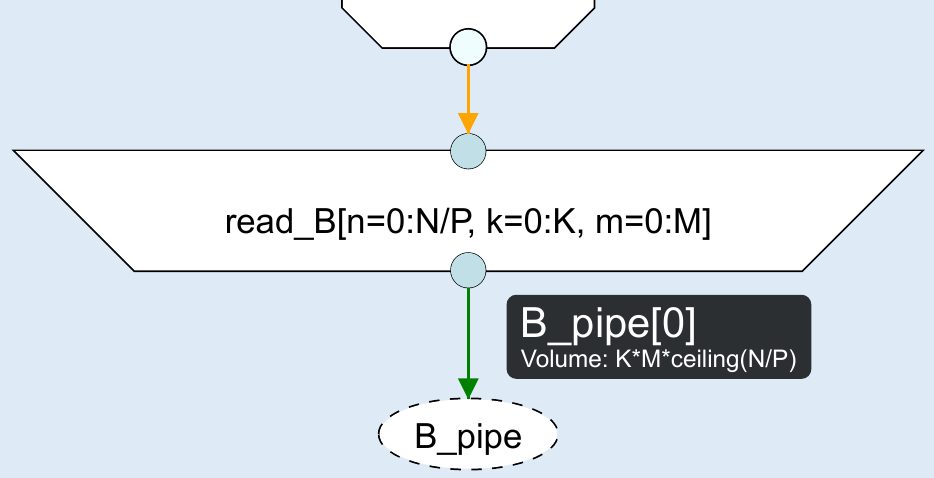}
    \end{minipage}
    \begin{minipage}{.49\columnwidth}
        \includegraphics[width=.99\columnwidth]{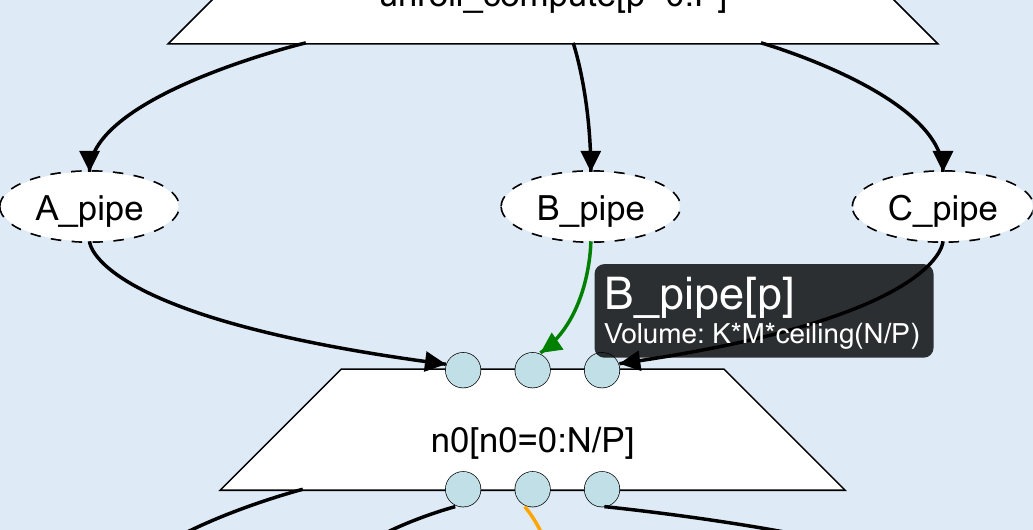}
    \end{minipage}
    \caption{Data movement volume annotations on the producer (left) and the consumer (right) of a stream, used to verify correctness of the program.}
    \label{fig:memlet_volume}
\end{figure}

Because all data movement is explicitly captured in the SDFG, programmers can benefit from the information annotated on dataflow edges to verify the correctness of producer/consumer relationships, which are automatically inferred by the tool based on the access pattern expressed by map scopes in the graph. Figure~\ref{fig:memlet_volume} shows the annotation of a dataflow edge written by the processing element prefetching the matrix \texttt{B} from Figure~\ref{fig:systolic_array} into the stream object \texttt{B\_pipe}, and the corresponding read from \texttt{B\_pipe} within the processing element. The matrix of size $K{\times}M$ is read $N/P$ times, where $P$ is the tile size introduced by the systolic array (see Section~\ref{sec:systolic_arrays}), resulting in a data volume of $K\cdot M \cdot \frac{N}{P}$ annotated on the dataflow edge/memlet.

\textbf{Code generation.} Due to the distinct methods of expressing processing elements, the semantics of allocating streams varies significantly between the Xilinx and Intel backends. When generating Xilinx code, streams are emitted in the top-level kernel function as local objects, where they must be passed as arguments to the producer and consumer accessing them (see Figure~\ref{fig:processing_element_xilinx}). For Intel OpenCL codes, they must be emitted to the global kernel scope, where the appropriate producer and consumer will read them directly (i.e., rather than receiving them as arguments).

\subsection{Parametric Processing Elements: Systolic Arrays}
\label{sec:systolic_arrays}

\textbf{Representation.} Systolic arrays~\cite{systolic_arrays} are a powerful pattern to express parametric parallelism through deep pipelines, and are the most potent source of parallelism on modern FPGAs~\cite{hls_transformations} when applicable. SDFGs expose this pattern 
through unrolled maps in the outermost FPGA kernel scope, with a parametric --- but compile-time specialized --- number of iterations, coupled with \emph{arrays} of stream objects. When such a map is unrolled, each instance semantically becomes a weakly connected component in the state, resulting in them being instantiated as separate processing elements according to the semantics in Section~\ref{sec:processing_elements}. This is equivalent to any other map construct in the SDFG: namely, they represent independently executable replications of the contained subgraph (unrolled maps can occur at any level of nesting in the program), but are recognized as a special case in the top-level scope of an FPGA kernel.

An SDFG implementing a one-dimensional systolic array for matrix multiplication ($\mathbf{C}=\mathbf{A}{\times}\mathbf{B}$) is shown in Figure~\ref{fig:systolic_array}, where the map nodes annotated by red borders instantiate the systolic array. Each element implements the same content (highlighted), but reads from a distinct index in three arrays of stream objects (``pipes'') for \texttt{A}, \texttt{B}, and \texttt{C}, respectively. Since every processing element is only connected to the previous and the next, they must pass data along the chain~\cite{gemm_hls} from the head towards the tail. The processing elements implement a simply buffering scheme where each element stores one element of \texttt{A} in a local buffer, then streams over the full \texttt{B} matrix, before writing back a complete output tile of \texttt{C}. This simple SDFG already yields $364$ and $\SI{188}{\giga\op\per\second}$ with $8k{\times}8k$ matrices when compiled for an Intel Stratix~10 and a Xilinx Alveo~U250 board, respectively, with much potential for additional optimization~\cite{gemm_hls}.

\textbf{Code generation.} Systolic array code generation varies between vendors due to the different ways of expressing processing elements. In Xilinx codes, it is sufficient to unroll a loop in the C++ kernel code with bounds known at compile time, letting constant propagation fix all the indices in each instantiation to lay out the systolic array, as shown in Figure~\ref{fig:processing_element_xilinx}. For Intel, the OpenCL kernel itself is replicated and specialized directly in the code generator (see Figure~\ref{fig:processing_element_intel}).

\subsection{Memory Hierarchy}
\label{sec:memory_hierarchy}

\textbf{Representation.} Not all data movement is born equal: dataflow can have significantly different performance impact depending on the location and storage type of the source and destination, even though the number of bytes moved are the same. The FPGA backend exposes \emph{global} memory, which represents data present in off-chip, memory-mapped storage such as DDR or HBM; \emph{local} memory, representing any on-chip memory implementation such as registers, BRAM/M20K, LUTRAM, or UltraRAM (left up to the HLS compiler); \emph{registers}, which is a subset of local memory, but forces the HLS compiler to allow fully parallel read/write access to every entry of the container; and experimental support for \emph{shift registers}, implementing cyclic buffering patterns with multiple access points (natively supported by Intel OpenCL). Combining these allows implementing highly specialized memory hierarchies, as well as host/device interaction, in a way that is compatible with both Xilinx and Intel devices.

\textbf{Code generation.} Local memories can be emitted as regular C arrays directly in the kernel code, while off-chip memory is allocated with API calls in the host code and passed to the kernel arguments.
The FPGA backend gives the underlying HLS compilers additional scheduling freedom by generating a distinct pointer argument for every access to the same DRAM memory container present in the kernel and marking them as \texttt{restrict} in OpenCL, such that every read and write can be performed independently, which is safe due to SDFG semantics.

Whenever both reads and writes are emitted to local memory, and the write is not marked as a potential conflict, the generated code is annotated with pragmas to instruct the compiler to ignore dependencies (\texttt{HLS DEPENDENCE} for Xilinx and \texttt{ivdep} for Intel). This is implied by SDFG semantics, where these accesses are either in dataflow sections (where conflicts must be annotated), or in control flow scopes, that are inherently sequentialized.

\section{Multi-Level Design with Library Nodes}
\label{sec:library_nodes}
\label{sec:multi_level}

\begin{figure}
    \centering
    \includegraphics[width=.7\columnwidth]{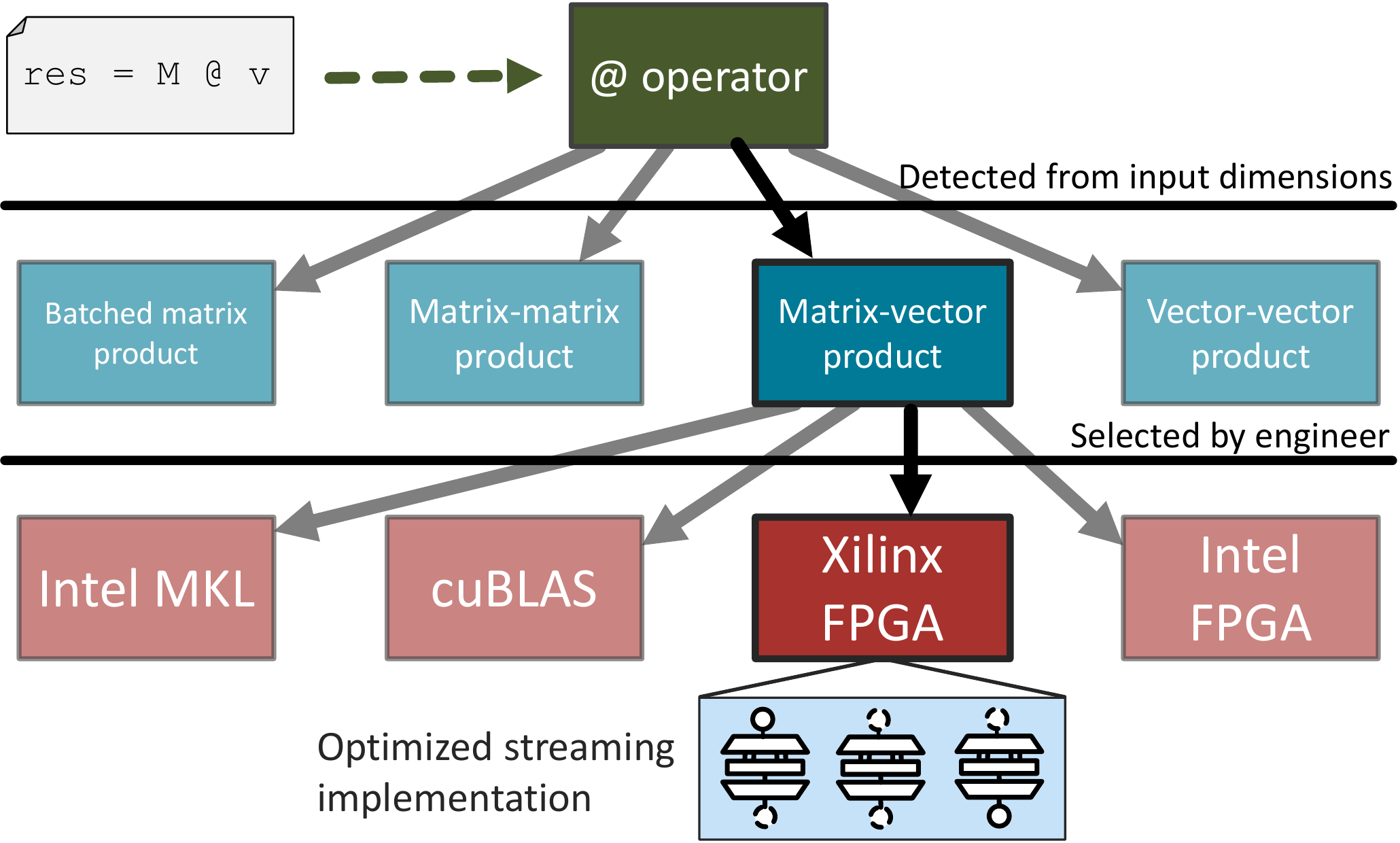}
    \caption{Multiple levels of nested Library Node expansions.}
    \label{fig:nested_library_nodes}
\end{figure}

\begin{figure*}
    \centering
    \begin{minipage}[b]{.39\textwidth}
        \centering
        \begin{minted}{Python}
n = dace.symbol("n")
a = dace.symbol("a", dtype)

@dace.program
def axpydot(x: dtype[n], y: dtype[n],
            w: dtype[n], result: dtype[1]):
    z = np.ndarray([n], x.dtype)
    blas.Axpy(a, x, y, z)
    blas.Dot(z, w, result)
        \end{minted}
        \caption{Implementation of \texttt{AXPYDOT} using the standard DaCe Python frontend and BLAS library calls.}
        \label{fig:fblas_frontend}
    \end{minipage}\hfill%
    \begin{minipage}[b]{.19\textwidth}
        \centering
         \includegraphics[height=.11\textheight]{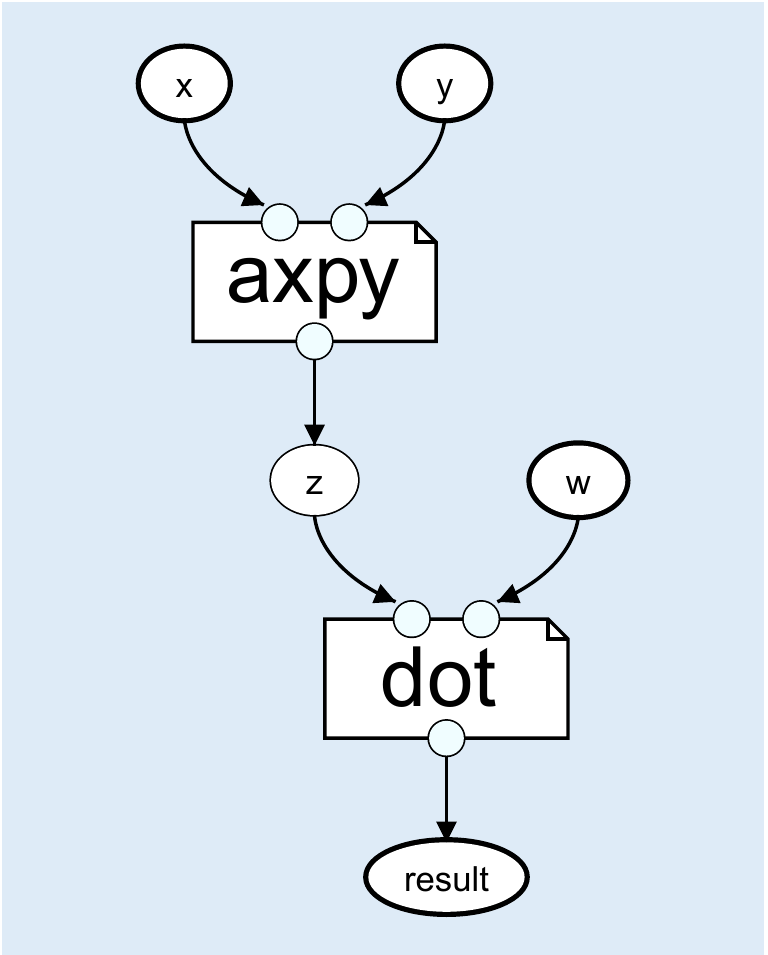}
        \caption{Generic SDFG computing \texttt{AXPYDOT}.}
        \label{fig:fblas_sdfg_initial}
    \end{minipage}\hfill%
    \begin{minipage}[b]{.39\textwidth}
        \centering
        \includegraphics[height=.11\textheight]{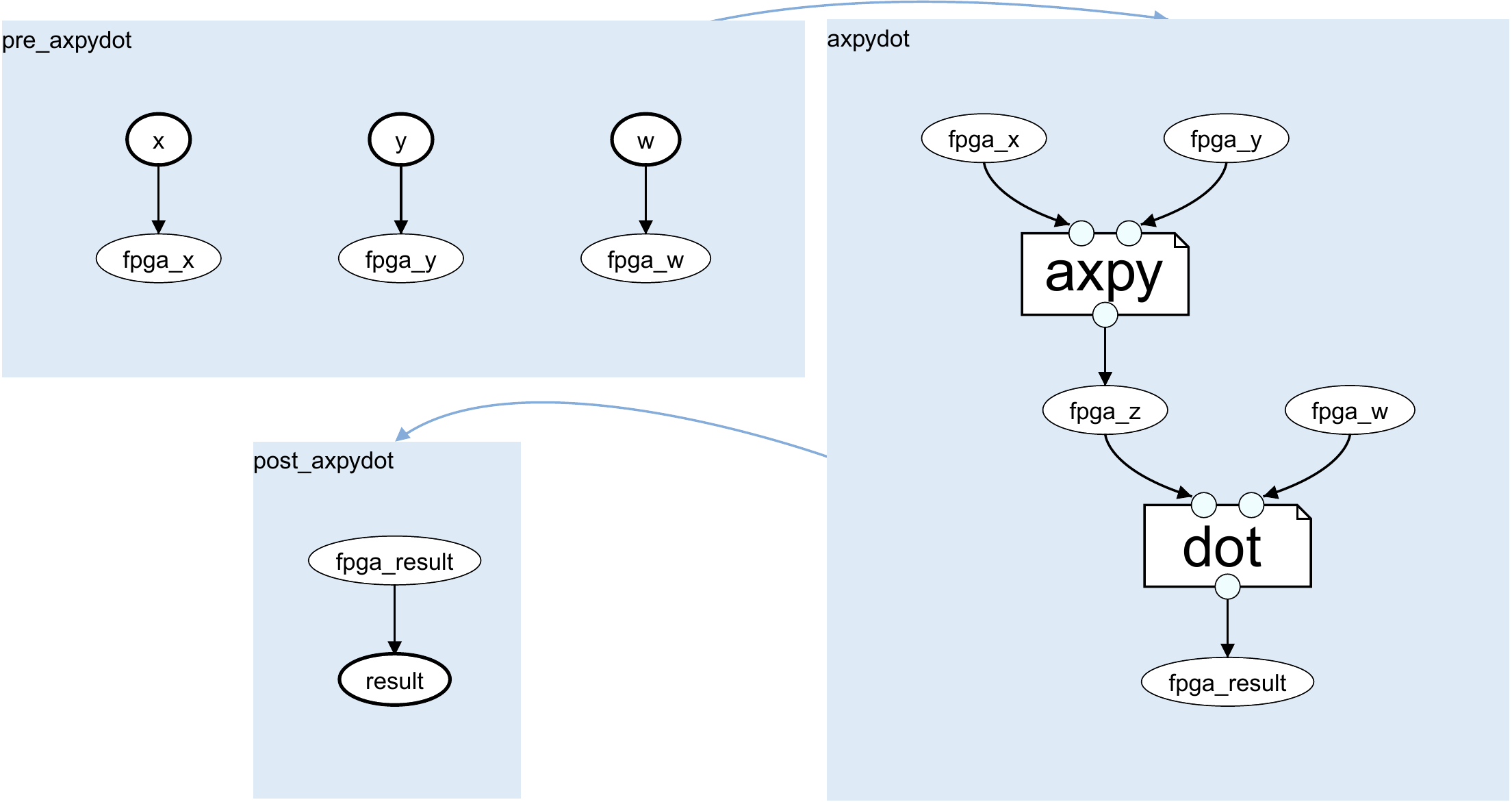}
        \caption{The SDFG from Figure~\ref{fig:fblas_sdfg_initial} automatically transformed for FPGA execution.}
        \label{fig:fblas_sdfg_fpga_transform}
    \end{minipage}
\end{figure*}


SDFGs provide a data-centric view of the implemented application, enabling a wide range of optimization opportunities~\cite{dace}, which can be exploited using graph-rewriting transformations on the SDFG.
Some optimizations, however, arise from knowledge about the underlying application domain (for example, algebraic identities), which are difficult or impossible to express generally without encoding domain-specific knowledge into the representation.
To accommodate such domain-specific optimizations, we will use Library~Nodes to represent an \emph{abstract behavior} (the ``what'') on the incoming/outgoing data connectors (as opposed to a concrete implementation of this behavior, the ``how'').
Library Nodes are \emph{expanded} by replacing them with a subgraph, ``lowering'' them towards a concrete implementation of their behavior.
During this expansion, Library Nodes can inspect their context using the surrounding memlets and nodes, which may change the structure of the expanded subgraph, e.g., by checking if inputs or outputs are streams or if the use vector types.

For example, in a neural network program, a Library Node can represent a convolution applied to a given input format and producing a given output format. However, \emph{which} convolution is applied, and how/where it is executed, can be deferred, exploiting the high level of abstraction to detect and apply domain-specific transformations. Before an SDFG can be used to generate code, all Library Nodes must be fully expanded to native SDFG constructs, but can go through several levels of lowering~\cite{mlir} before reaching a fully expanded state.

Domain-specific transformations are not the only opportunity enabled by the Library Node abstraction. 
The ability to \emph{nest} levels of decreasing abstraction make Library Nodes a versatile tool to achieve a number of tasks. An example is included in Figure~\ref{fig:nested_library_nodes}: a linear algebra-aware frontend produces an SDFG containing a generic matrix multiplication operator Library Node (top); once expanded, the Library Node can delegate itself to a number of different linear algebra operations, depending on the dimensionality of the two operands; finally, the performance engineer can choose either a generic implementation, or one specialized for a specific backend. For example, they may choose an FPGA-specific streaming implementation specialized for Xilinx FPGAs (or a generic FPGA implementation).

In the following, we use two composite linear algebra kernels from the extended set of BLAS subprograms proposed by Blackford~et~al.~\cite{extended_blas} to demonstrate the multi-level design process using SDFGs and the DaCe framework.

\subsection{High-Level Domain-Specific Frontends}
\label{sec:domain_specific_frontends}

To develop programs using the SDFG representation, programmers can use high-level frontends, rather than using the low-level graph API to create SDFGs directly. For example, the DaCe framework itself exposes a Python frontend supporting NumPy \cite{p4}, and with BLAS extensions. Calling BLAS routines or using linear algebra operators on NumPy arrays will emit BLAS Library Nodes in the resulting SDFG, which can later be expanded to the desired implementation: e.g., direct function calls to MKL, cuBLAS, or OpenBLAS, or specialized SDFG subgraphs targeting a specific architecture.

Figure~\ref{fig:fblas_frontend} shows \texttt{AXPYDOT}, a small composite BLAS kernel, summing two input vectors, then taking the dot product with the resulting vector and a third input vector.
The SDFG emitted by the frontend for this code is shown in Figure~\ref{fig:fblas_sdfg_initial}. The BLAS operators are instantiated as the \texttt{axpy} and \texttt{dot} Library Nodes, reading and writing from arrays. The two kernels exchange data through the array \texttt{z}, which will be first written by \texttt{axpy} and then read by \texttt{dot}, in sequence.
Kernels that are composed of BLAS level 1 and 2 routines, such as \texttt{AXPYDOT}, are fully memory bound, but expose a promising opportunity for streaming computation by pipelining temporaries directly between subroutines~\cite{fblas} on the FPGA, which we will exploit in the following.

\subsection{Mid-Level FPGA Transformations}
\label{sec:transformations}

SDFGs can be specifically engineered to target FPGAs, by writing them using the graph API. Alternatively, DSLs that target FPGAs, such as StencilFlow~\cite{stencilflow}, can directly emit FPGA-specific graphs. Finally, existing SDFGs can be transformed from a generic implementation to an FPGA implementation using graph transformations. Any of these approaches will result in graphs that can be further optimized using general-purpose transformations available in the DaCe toolbox. This includes platform-agnostic transformations such as map tiling, inserting fast memory buffers, or removing redundant memory accesses~\cite{dace}; and more FPGA-oriented transformations, which we describe here.

\subsubsection{Transforming a Subgraph into an FPGA Kernel}
\label{sec:fpga_transform}

If the input is a generic graph that has not yet been targeted to FPGAs, programmers can automatically offload a full SDFG or a specific subgraph for FPGA execution using the \texttt{FPGATransformSDFG} and \texttt{FPGATransformState} transformations, respectively, provided in the DaCe framework. These detect all DRAM accesses in the target graph or subgraph, then create additional pre- and post-states performing memory transfers between host and device. The memories accessed by the transformed subgraph are replaced with their FPGA equivalents.

Figure~\ref{fig:fblas_sdfg_fpga_transform} shows the \texttt{AXPYDOT} example from Figure~\ref{fig:fblas_sdfg_initial} after applying the \texttt{FPGATransformSDFG} transformation. Occurrences of the DRAM memories \texttt{x}, \texttt{y}, and \texttt{w} and replaced with corresponding FPGA memories \texttt{fpga\_x}, \texttt{fpga\_y}, and \texttt{fpga\_w} in the kernel graph, the memories are copied to the FPGA before the kernel is executed in the state \texttt{pre\_axpy}, and the output array \texttt{result} is copied back in the state \texttt{post\_axpy}. This program can already be generated and compiled for both Xilinx and Intel boards.

\subsubsection{Memory Access Extraction}
\label{sec:memory_extraction}

\begin{figure}
\end{figure}

When the memory access pattern of a certain computation is known, it is often beneficial to stream the data into the FPGA processing elements. Creating streaming accessors has many benefits~\cite{hls_transformations}, including making use of burst-mode in memory controllers, tailored buffering for pipelined execution, broadcasting off-chip memory to multiple processing elements, or customizing caching mechanisms. 

In DaCe, extracting a streaming pattern from an existing memory access to a streaming access is performed via the \texttt{StreamingMemory} transformation. The transformation processes the outgoing (or incoming) memlets of a certain data access node, finding all recurring access patterns of unique symbolic expressions. If the range consists of one scalar or vector element, the transformation can be applied. It then extracts the read (write) out of the computation by introducing another component that accesses the memory in the same order as the computation, and pushes it onto a stream (or pops computation outputs and stores results). The corresponding outgoing/incoming memlets are replaced by memlets that access the stream instead.
If more than one PE uses the same memory access order, the transformation generates a single streaming component that connects one array node to multiple streams. In order to avoid deadlocks, the transformation also detects dependencies by computing reachability from the destinations/sources of the memlet paths (inherently given by the construction of the SDFG). If accesses are dependent, separate components are created, even for the same access pattern.

\subsubsection{Pipeline Fusion} 
\label{sec:pipeline_fusion}

Following streaming the endpoints of a computation, we also consider streaming composition of consecutive computations. In unoptimized SDFGs, intermediate data is represented as data access nodes, pointing to off-chip memory by default. This round-trip is undesirable, and in certain computations can be completely avoided by fusing the two underlying pipelines into one.

The \texttt{StreamingComposition} transformation is similar in structure to memory access extraction, but checks for array nodes with in-degree and out-degree of one, which contain equivalent access orders that can be composed. To do so, the transformation traces the memlet path through map/pipeline scopes and nested SDFGs, canonicalizing the memlets' symbolic expressions by remapping symbol names to indices. If the ranges of the iteration spaces match exactly, and the symbolic expressions are equal, the result of the first computation can be streamed into the second.
Similarly to \texttt{StreamingMemory}, we replace the memory access nodes and neighboring memlets with streams, converting global memory arrays into local streams.

\begin{figure}[t]
    \centering
    \includegraphics[width=.99\columnwidth]{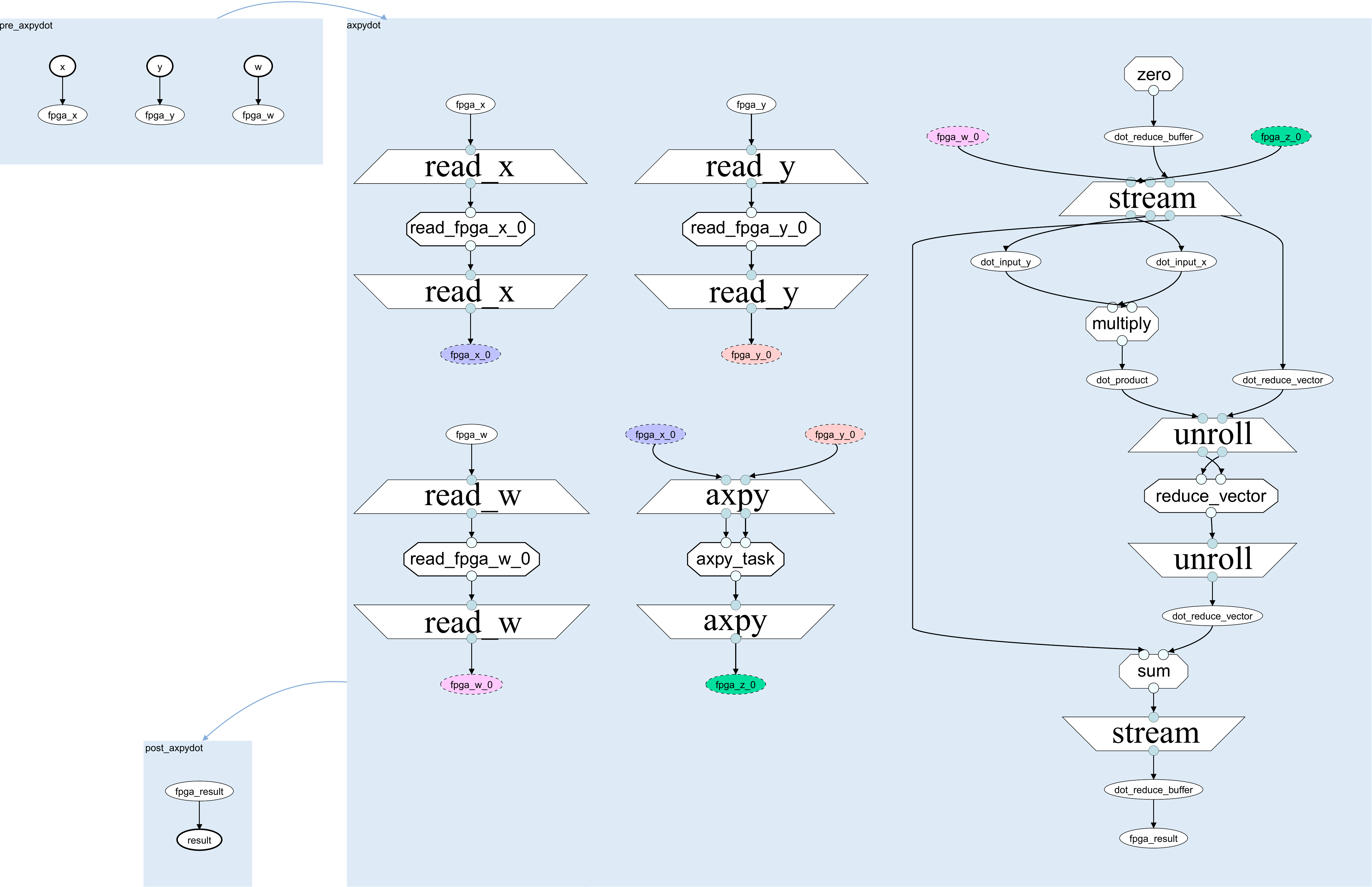}
    \caption{The \texttt{AXPYDOT} program after automatically extracting memory accesses into processing elements and streaming between operators. Streams are color-coded by name for pipeline visualization.}
    \label{fig:fblas_sdfg_fpga_streaming}
\end{figure}

\subsubsection{Putting it All Together} 

An example that uses all mid-level transformations 
can be seen in Figure~\ref{fig:fblas_sdfg_fpga_streaming}. Manually applying \texttt{FPGATransformSDFG}, \texttt{StreamingMemory} on $x,y,w$ and \texttt{StreamingComposition} on \texttt{AXPYDOT} yields fully pipelined execution. The same transformations can be applied automatically, in a scheme which we perform for our applications in Sections \ref{sec:case:la}--\ref{sec:case:sf}. However, the transformations must be attempted at a certain order in order to apply. 
First, the input SDFG must be transformed to FPGA-based computation (\texttt{FPGATransformSDFG}). Then, the data can be vectorized to the desired length (using \texttt{Vectorization}), which the Library Nodes use to control unrolling and accumulation factors upon expansion. After expanding the Library Nodes, the memory access patterns of each computation is exposed, and \texttt{StreamingMemory} and \texttt{StreamingComposition} can be applied.
Lastly, the memory assignment to banks can be tweaked by inspecting the dataflow of the SDFG.

The resulting SDFG is capable of generating separate modules for efficiently reading/writing off-chip memory, using the stream construct as a FIFO queue to connect pipeline stages. In the rest of this section, we describe how the Library Nodes and can be further specialized depending to the target FPGA vendor.

\subsection{Platform Specialization}
\label{sec:platform_specialization}

While it is possible (and often sufficient) to implement operators with a generic SDFG subgraph, it can be desirable to specialize the implementation of an operation to a specific target. This can simply mean calling an external high performance implementation, such as cuBLAS, or it can mean a specialized native DaCe graph expansion. Because of the differences in capabilities between Intel and Xilinx architectures, such specializations prove useful in practice.

\subsubsection{Floating Point Accumulation}
\label{sec:accumulation}

For computations that need to perform accumulation, such as the \texttt{DOT} operator used in Figure~\ref{fig:fblas_sdfg_initial}, it is beneficial to specialize the computation based on whether the underlying architecture supports accumulation on the given data type.
Intel Arria~10 and Stratix~10 architectures supports native 32-bit floating point accumulation, which allows a stream of floats to be directly summed into an output register.
Contemporary Xilinx FPGAs, such as the Alveo U250, do not have native 32-bit floating point units, and cannot directly accumulate floating point numbers into a single register, as this results in a loop-carried dependency induced by the multiple-cycle latency of the addition operation.
For 64-bit floating point, neither Xilinx nor Intel support accumulation, and both must address the issue of loop-carried dependencies.

\begin{figure}
    \centering
    \begin{minipage}[b]{.48\columnwidth}
        \centering
        \includegraphics[width=.915\columnwidth]{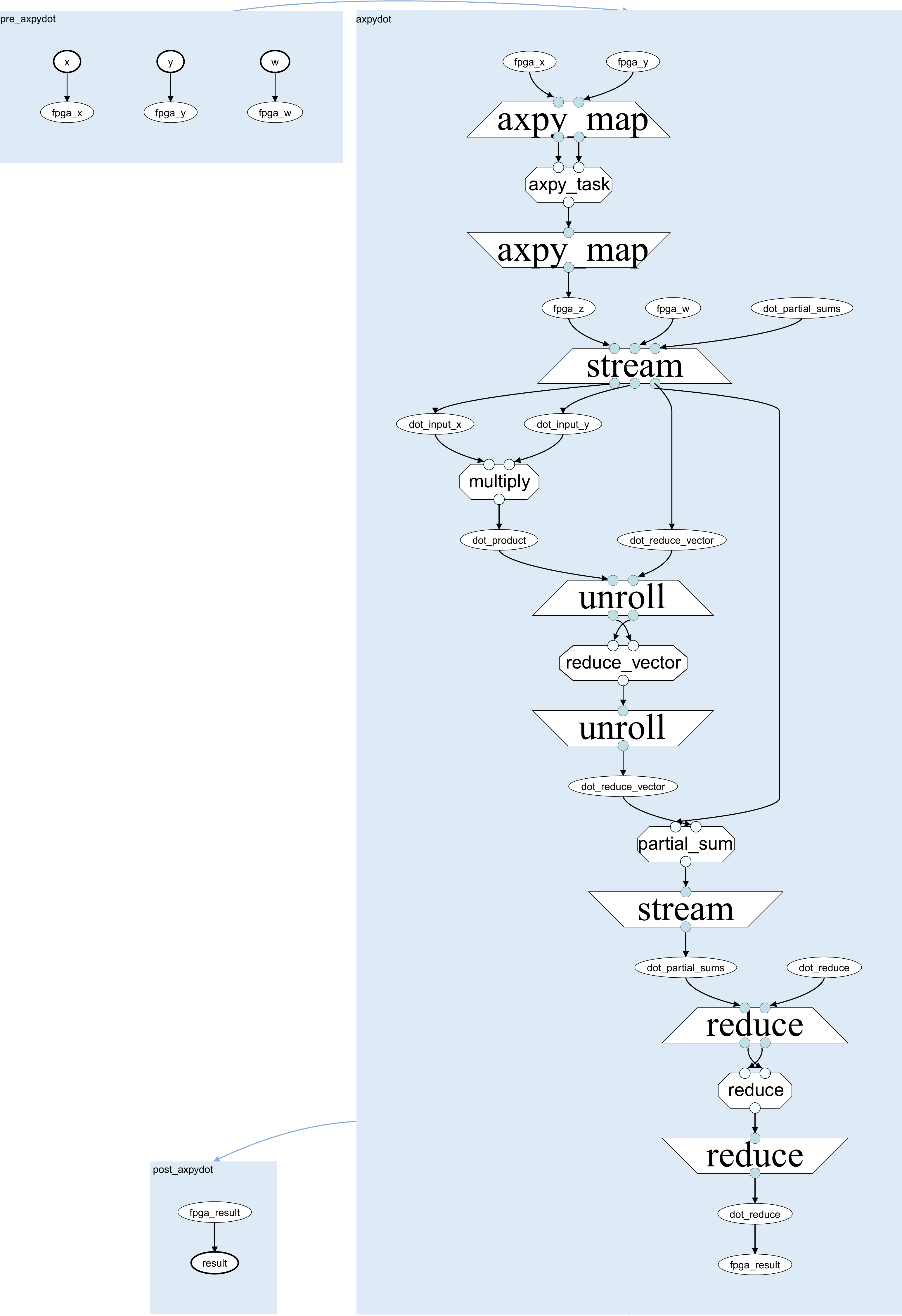}
    \end{minipage}\hfill%
    \begin{minipage}[b]{.48\columnwidth}
        \centering
        \includegraphics[width=.915\columnwidth]{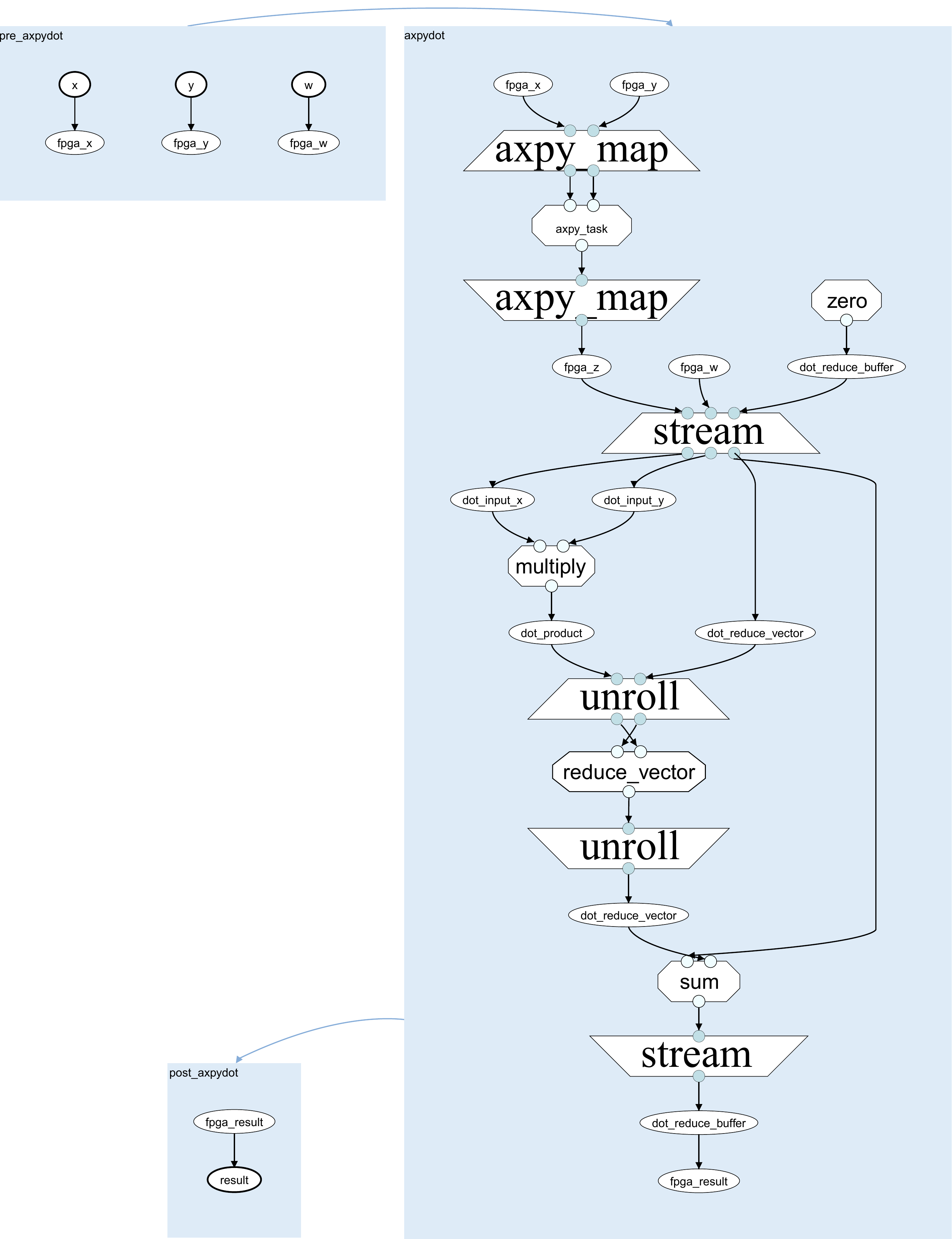}
    \end{minipage}
    \caption{\texttt{AXPYDOT} expanded for Xilinx (left), using a partial sum and reduce phase, and for Intel (right), accumulating into a single register.}
    \label{fig:fblas_sdfg_expanded}
\end{figure}

To avoid the loop-carried dependency for \texttt{DOT}, we can perform \emph{accumulation interleaving}~\cite{hls_transformations} by summing the incoming data into a number of \emph{partial} sums, stored in a buffer of a size larger than the latency of the addition operation. In Figure~\ref{fig:fblas_sdfg_expanded}, the \texttt{AXPY} and \texttt{DOT} operators have both been expanded for Xilinx (left) and Intel (right). \texttt{AXPY} uses a generic implementation (identical to the CPU implementation), while \texttt{DOT} is implemented using specialized expansions depending on the target architecture. The Xilinx-targeted expansion uses a partial sum strategy to resolve the loop-carried dependency, using two unrolled maps. The first (``\texttt{unroll}'') sums up all entries of the vector containing the product of contributions from $x$ and $y$ using a fully unrolled circuit (i.e., $W{-}1$ adders, where $W$ is the vectorization width), resulting in a single element contribution. This contribution is added into the partial sum buffer, accessed with a cyclic index. The second unrolled map (``\texttt{reduce}'') is performed after the main streaming phase, and sums up all values in the partial sums buffer into a single output, which is written to the output (again consuming $W{-}1$ adders. In a resource-constrained scenario, this could be reduced to a single adder without impacting the asymptotic runtime). The Intel specialization instead accumulates into a single register, saving the partial buffer and additional reduction.

\subsubsection{Shift Registers}
\label{sec:shift_registers}

For Intel FPGA, we can use shift registers to achieve the sliding window-style buffering pattern required for stencil computations. This is a powerful abstraction that is desirable to exploit by the Intel FPGA backend, but does not (as of writing) exist in Vivado~HLS. In Section~\ref{sec:stencilflow_expansions}, we show how --- with some additional effort --- the shift register pattern can be imitated for stencil computations on Xilinx, by implementing a Xilinx-specific expansion using explicit buffers between each access point.

\subsubsection{Specializing SDFGs vs. Hand-Written Code}

Targeting low-level aspects of the HLS tools, such as floating point accumulation and shift registers, raises the question of whether there is even any benefit of using the Library Node expansion abstraction versus writing these specializations by hand. While embedding hand-written HLS code inside SDFGs is indeed supported natively by the framework, implementing the specializations as graph expansions come with a significant benefit of \emph{malleability}, \emph{analyzability}, and potential for reuse. \emph{Malleability}, because the expanded subgraph can be further transformed by the DaCe framework: for example, to tile the internal maps, or to add/change sizes and constants, such as input size and vectorization width. \emph{Analyzability}, because the graph view gives an overview of the computational structure, and allows inspecting data volumes on dataflow edges (e.g., to detect producer/consumer mismatches). Potential for reuse, because even specialized codes can potentially be reused between platforms or vendors -- for example, using the partial sums implementation of \texttt{DOT} to reduce 64-bit floating point numbers on an Intel platform, without writing any additional code. Finally, staying within the SDFG representation lets the program benefit from the powerful code generating backends, which are continuously improved and updated to work seamlessly with the newest versions of the HLS toolflows.

\section{Case Study: Linear Algebra} \label{sec:case:la}

We have seen the multi-level design methodology applied to the \texttt{AXPYDOT} example throughout Section~\ref{sec:multi_level}. We additionally show how this flow can reproduce the composite BLAS kernel \texttt{GEMVER} evaluated by \fblas{}~\cite{fblas}, using our multi-level SDFG design.
In the following, we evaluate kernels on a Xilinx Alveo U250 accelerator board, compiled using Vitis~2020.2 for the \texttt{xilinx\_u250\_xdma\_201830\_2} shell, and using the Xilinx Runtime (XRT) version 2.5.309. 
Results were measured 10 times, the median and 95\% nonparametric confidence intervals are listed as errors.

\subsection{\texttt{AXPYDOT} Evaluation}

\begin{table}[t]
    \centering
    \caption{Performance of \texttt{AXPYDOT} on the Alveo U250 (Xilinx) FPGA.}
    \begin{tabular}{c c}
    \toprule
    \bf Na\"{i}ve HLS in DaCe & \bf Streaming Transformations \\\midrule
    $3.57\pm 0.15$ GB/s     & $9.34\pm 0.03$ GB/s\\
    \bottomrule
    \end{tabular}
    \label{tab:axpydot}
\end{table}

The result of the \texttt{AXPY/DOT} BLAS operation composition are listed in Table \ref{tab:axpydot}. 
As the program is bandwidth-bound, we list the attained bandwidth of \texttt{AXPYDOT} when run with an input buffer of 800 MiB (209,715,200 elements).
The table shows that the streaming transformations, which are applied automatically, are able to stream all memory accesses, and fuse the \texttt{AXPY} and \texttt{DOT} pipelines.
This is a promising result for FPGA programs in general, since the transformations detected the access patterns directly and did not employ application-specific knowledge.
As for generated code length, the na\"{i}ve version generates one module (processing element) and 139 lines of code, whereas the streamed version generates 5 separate modules and 207 lines of code.
All in all, this contributes to a 2.6$\times$ speedup of the streamed version over the original input graph.

\subsection{\texttt{GEMVER} Optimization and Evaluation}

\begin{wrapfigure}{r}{0.4\linewidth}
\begin{center}
    \includegraphics[width=.9\linewidth]{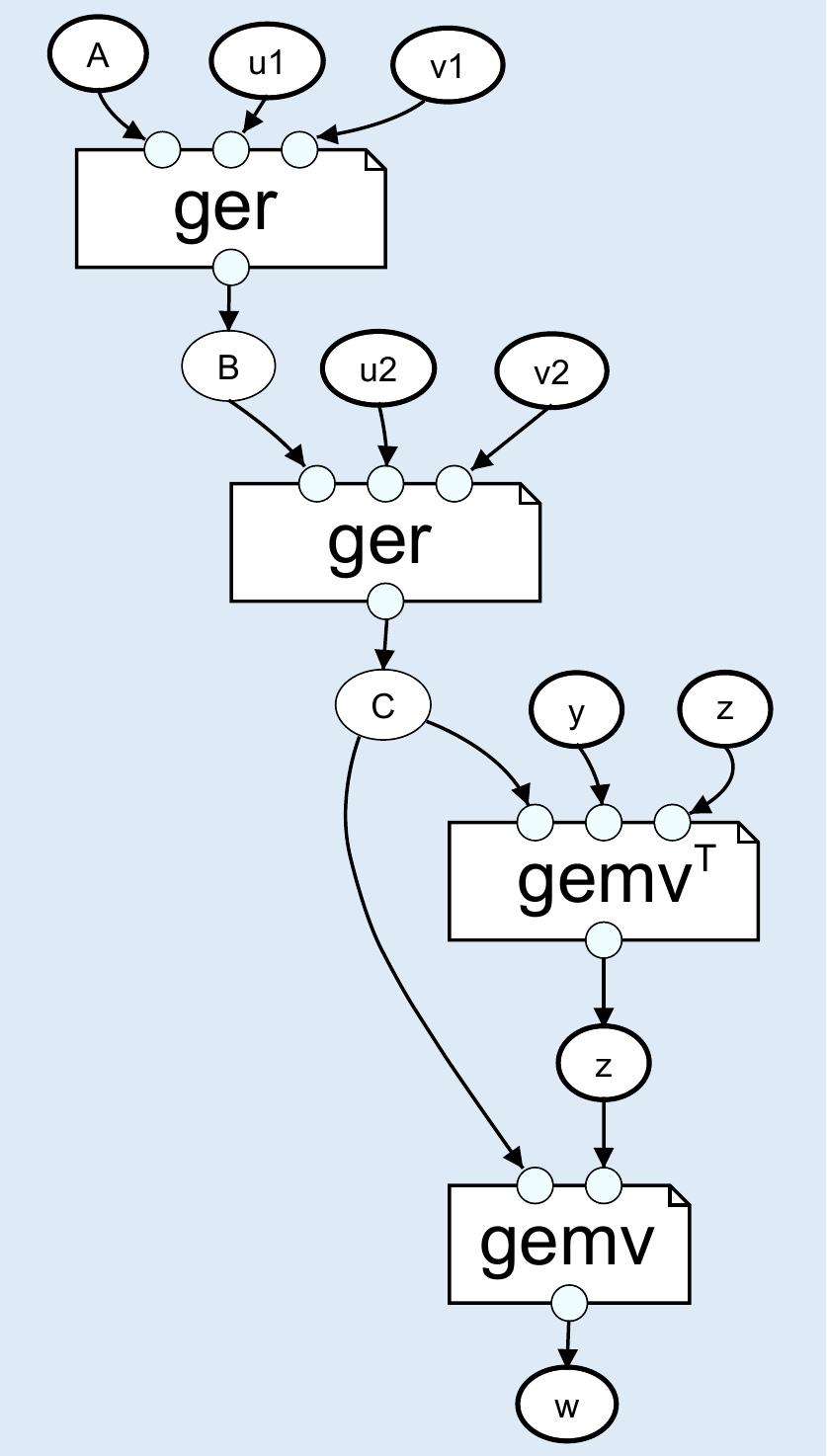}
    \caption{\texttt{GEMVER} SDFG.}
    \label{fig:gemver}
\end{center}
\end{wrapfigure}

The \texttt{GEMVER} application is a composition of several BLAS operations used in solving systems of equations. Its SDFG is shown in Figure~\ref{fig:gemver}. Specifically, \texttt{GEMVER} performs two rank-1 updates (\texttt{GER}), a transposed matrix-vector multiplication (\texttt{GEMV}$^\text{T}$), and another matrix-vector multiplication (\texttt{GEMV}), both using different access patterns.

There are several approaches that can be taken to layout this complex composition~\cite{fblas}, stemming from the difference in access patterns between the two \texttt{GEMV}s, and the inability to directly stream the result of the two rank-1 updates into both matrix-vector multiplications at the same time. Thus, malleability plays a key role for \texttt{GEMVER} optimization.

In DaCe, Library Nodes can be expanded in multiple ways, depending on the target platform and the parameters of the node (i.e., whether an array is transposed). For \texttt{GER}, vectorized expansion would only apply if the matrix and the first or second vector are vectorized, but not both. In order to compose the operators \texttt{GEMVER}, the performance engineer must match the tiling schemes between them: for the transposed \texttt{GEMV}, a scheme that streams in tiles by columns matches the output of \texttt{GER}, while the second \texttt{GEMV} can use a row-major scheme. Once the access patterns match, the array \texttt{z} can automatically be streamed by the mid-level transformations.




\begin{table}[b]
    \centering
    \caption{Performance of \texttt{GEMVER} on the Alveo U250 (Xilinx) FPGA.}
    \begin{tabular}{l l l}
    \toprule
    \bf Version & \bf Runtime [s] & \bf Off-Chip Volume \\\midrule
    
    Na\"{i}ve SDFG & 1.10$\pm$0.029 & 6.0 GiB (---) \\
    Manual memory banks & 1.52$\pm$0.000 & 6.0 GiB (1$\times$) \\
    Streaming composition & 0.88$\pm$0.004 & 4.0 GiB (1.5$\times$) \\
    Manual composition & \textbf{0.74}$\pm$0.005 & 3.0 GiB (2$\times$) \\
    \bottomrule
    \end{tabular}
    \label{tab:gemver}
\end{table}

Data movement for the two consumers of $C$ can be significantly reduced by \textit{both} pipelining \textit{and} storing it in off-chip memory for later use. As mentioned in Section~\ref{sec:pipeline_fusion}, streaming composition only works if there are no other uses of the array. However, the performance engineer can manually replicate $C$ (interactively or programmatically) following Library Node expansion, creating the possibility to apply pipeline fusion once more, which would remove one of the replicas of $C$ in favor of a stream.

Results of \texttt{GEMVER} with $N$=16,384 are shown in Table \ref{tab:gemver}.
As our baseline already uses vector width of 16 elements, tiled computation, and the matching Library Node implementations, the differences in performance are not substantial.
The example shows the importance of streaming composition: when memory banks are manually chosen to maximize bandwidth the program is slower. Performance only improves when the accesses are streamed and composed.
Using the manual replication of $C$ yields an additional improvement in performance: a 2$\times$ reduction in memory movement and 1.49$\times$ reduction in overall runtime.

\section{Case Study: Machine Learning} \label{sec:case:ml}

We present the multi-level design methodology for FPGAs in the context of a machine learning application, utilizing the ONNX frontend of DaCe.

\subsection{ONNX Domain-Specific Frontend}
DaCeML\footnote{\url{https://github.com/spcl/daceml}} is a data-centric machine learning framework~\cite{daceml} based on DaCe. The framework exposes the operators of the Open Neural Network eXchange (ONNX) IR as Library Nodes, and provides a collection of implementations for each operator, specialized for CPUs, GPUs, and FPGAs. 
%
\begin{figure}
 \begin{minipage}[b]{.69\columnwidth}
        \centering
        \begin{minted}[breaklines]{Python}
import torch.nn as nn
import torch.nn.functional as F

@daceml.pytorch.dace_module
class LeNet(nn.Module):
  def __init__(self):
    self.conv1 = nn.Conv2d(1, 6, 5)
    self.conv2 = nn.Conv2d(6, 16, 5)
    self.fc1 = nn.Linear(256, 120)
    self.fc2 = nn.Linear(120, 84)
    self.fc3 = nn.Linear(84, 10)

  def forward(self, x):
    x = F.max_pool2d(
          F.relu(self.conv1(x)), 2)
    x = F.max_pool2d(
          F.relu(self.conv2(x)), 2)
    x = x.view(-1, 256)
    x = F.relu(self.fc1(x))
    x = F.relu(self.fc2(x))
    x = self.fc3(x)
    x = F.softmax(x, dim=1)
    return x
        \end{minted}
        \label{fig:lenet_model}
    \end{minipage}\hfill%
    \begin{minipage}[b]{.3\columnwidth}
        \centering
        \includegraphics[height=.275\textheight]{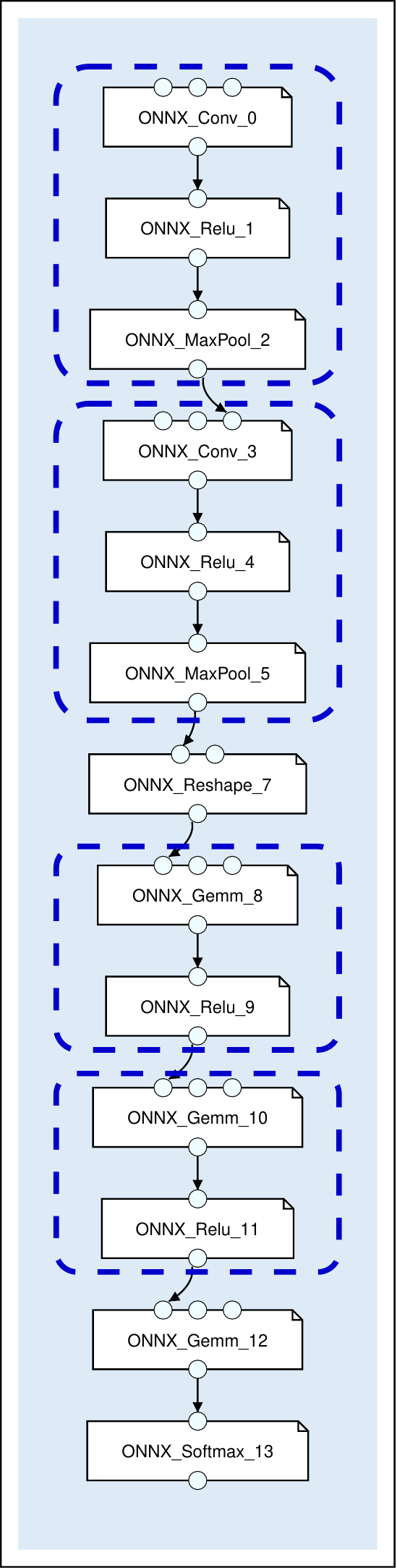}
    
    \end{minipage}
        \caption{LeNet-5: PyTorch module (left) and corresponding SDFG (right, array nodes are hidden for brevity).}
    \label{fig:lenet}
\end{figure}
As a case study, we present the process of lowering a PyTorch~\cite{pytorch} model, implementing Deep Learning inference with the LeNet-5 convolutional model~\cite{lenet}, to an SDFG that can be executed on either FPGA vendors. Using DaCeML, we generate ONNX files and SDFGs from PyTorch neural network modules with a single-line Python decorator (Figure~\ref{fig:lenet}).

DaCeML includes domain-specific transformations that are difficult or impossible to implement at the lower levels of the SDFG IR. To optimize for FPGAs, we develop and employ the \texttt{InputToConstant} transformation. DaCeML SDFGs typically receive both their inputs and parameters as arrays. For inference, those parameters can be fixed in hardware. This FPGA specific transformation converts a parameter array of the model to a compile-time constant. The transformation first verifies that the parameter array is never written to, then removes the input by traversing the edges, removing all corresponding memlets and access nodes. This domain-specific transformation requires knowledge of the parameter values, which are obtained from PyTorch. 


\subsection{Lowering to FPGA}

\begin{figure}[b]
    \centering
    \includegraphics[width=.9\columnwidth]{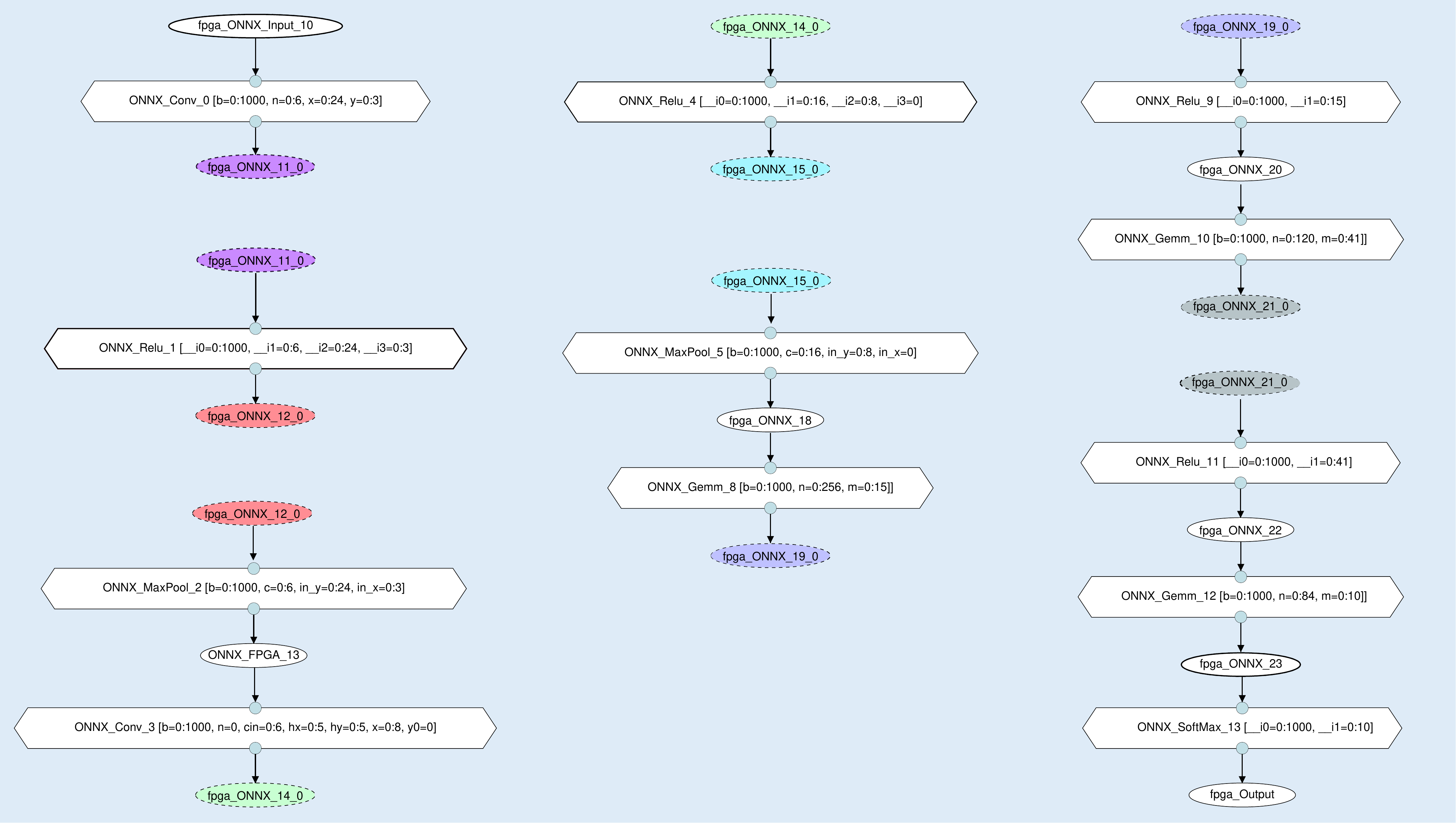}
    \caption{The LeNet program after transforming to stream between operators. Streams are color-coded by name for pipeline visualization. Library Node expansions are collapsed for readability.}
    \label{fig:lenet_sdfg_fpga_streaming}
\end{figure}

After applying the \texttt{FPGATransformSDFG} (see Section~\ref{sec:fpga_transform}) for offloading the execution of the SDFG on FPGA, each of the Library Node is expanded to a nested SDFG optimized for spatial architectures.
These expansions employ a range of optimizations. Convolutions (``\texttt{Conv}'') are implemented by using the image to column (\textit{im2col}) approach~\cite{im2col}. 
Therefore, the convolution and \texttt{GEMM} expansions rely heavily on the systolic matrix multiplication shown in Section~\ref{sec:systolic_arrays}.
The activation function (``\texttt{ReLU}'') is an element-wise operation, and can be expressed by nested maps. The \texttt{MaxPool} implementation uses a sliding window, implemented using shift registers.
All the network operators operate on single precision floating point, and can accept data either from off-chip memory or streamed in from on-chip memory.
The latter opens up to the possibility to streaming among operators, such as between convolution, activation and sub-sampling (blue dashed boxes in Figure~\ref{fig:lenet}), allowing a reduction of I/O through pipelined execution of sections of the graph. Figure~\ref{fig:lenet_sdfg_fpga_streaming} show the program SDFG obtained by applying the general \texttt{StreamingComposition} transformation (described in Section~\ref{sec:pipeline_fusion} to automatically achieve this).

\subsection{Evaluation}

For the evaluation, we target the BittWare 520N PCIe attached board,
with an Intel GX 2800 Stratix 10 processor, equipped with $4{\times}$ DDR4 memory
banks. The OpenCL code generated from DaCe is compiled
with version 20.3 of the Intel FPGA OpenCL SDK and 19.4 of the Quartus compiler.
We consider a batch size of $\SI{1000}{}$ for our experiments.
As reference, we use the inference time of running the input PyTorch model on CPU. We target a 36C/72T Intel Xeon Gold 6154 and PyTorch \texttt{v1.6.0}.
Inference is measured at $56.2\pm0.0433\:\si{\milli\second}$ and $14.4\pm0.0447\:\si{\milli\second}$ for 1 core and 36 cores, respectively.
We evaluate the different transformations applied to the SDFG in Table~\ref{tab:lenet}, showing a speedup of 1.87x over the single core CPU implementation. The \texttt{InputToConstant} transformation yields a speedup over the original SDFG graph of $3.2{\times}$ and a reduction of data movement. By applying the \texttt{StreamingComposition} to replace intermediate memories with streams, we further reduce the accessed data volume, and the speedup is increased to $8.8{\times}$. This shows how combining domain-specific and general transformations can yield significant benefits, and how general transformations are reused across domains (see examples for applying the transformation to linear algebra in Sections~\ref{sec:pipeline_fusion} and~\ref{sec:case:la}).

\begin{table}[ht]
    \centering
    \caption{Performance of \texttt{LeNet-5} on the Stratix 10 (Intel) FPGA with inference batch size 1000.}
    \begin{tabular}{l l l}
    \toprule
    \bf Version & \bf Runtime [ms] & \bf Off-Chip Volume \\\midrule
    
    Na\"{i}ve SDFG & 265.8$\pm$3.502 & 0.28 GiB (---) \\
    Input to constant & 81.3$\pm$1.570 & 0.22 GiB (1.2$\times$) \\
    Streaming composition & 30.1$\pm$0.703 & 0.16 GiB (1.7$\times$) \\
    \bottomrule
    \end{tabular}
    \label{tab:lenet}
\end{table}

\section{Case Study: StencilFlow} \label{sec:case:sf}

StencilFlow~\cite{stencilflow} is a domain-specific framework built on top of the DaCe framework, utilizing the full multi-level design to emit fully pipelined and deadlock free stencil architectures for complex input programs. In the following, we describe how each of the levels of optimizations described in Section~\ref{sec:multi_level} are applied in the context of StencilFlow, then further exploit the specialization capabilities of Library Nodes to extend the framework to target Xilinx FPGAs from the same input programs, even without access to shift registers.

\subsection{Stencil Language and Transformations}

StencilFlow defines a specialized stencil DSL, expressed in a JSON input format, allowing input programs with heterogeneous operators, reading from different input data containers, and with dependencies between them. An analysis tool parses the operators and maps the dependencies between them, computes the buffers required to fully pipeline each operator, then computes the delays between operators to insert \emph{delay buffers} between them to prevent deadlocks that can otherwise be induced by fork/joins in the dependency graph. An simple example program, implementing two iterations of the diffusion 2D stencil, is shown in Figure~\ref{fig:diffusion_json}.

\begin{figure}[b]
    \centering
\begin{minted}[breaklines, escapeinside=||]{json}
{"dimensions": [4096, 4096], "vectorization": 8,
  "outputs": ["d"], "inputs": {
  "a":  {"data_type": "float32", "input_dims": ["j","k"]},
  "c0": {"data_type": "float32", "input_dims": []},
  |...|
  "c4": {"data_type": "float32", "input_dims": []}},
 "program": {
  "b": {
   "data_type": "float32",
   "boundary": {"a": {"type": "constant", "value": 0}},
   "computation": "b = c0*a[j,k] + c1*a[j-1,k] + c2*a[j+1,k] + c3*a[j,k-1] + c4*a[j,k+1]"},
  "d": {
   "data_type": "float32",
   "boundary": {"b": {"type": "constant", "value": 0}},
   "computation": "c = c0*b[j,k] + c1*b[j-1,k] + c2*b[j+1,k] + c3*b[j,k-1] + c4*b[j,k+1]"
}}}
\end{minted}
    \caption{JSON-based StencilFlow program description.}
    \label{fig:diffusion_json}
\end{figure}

\subsection{Intel and Xilinx Stencil Specialization}
\label{sec:stencilflow_expansions}

StencilFlow presents results for an Intel Stratix~10 FPGA. The Intel OpenCL compiler is suitable to target stencil computations, due to the shift register abstraction allowing easy instantiation of more complex cyclic buffering patterns.
For this work, we show how the StencilFlow stack can be extended to target Xilinx FPGAs instead, simply by providing a new stencil node expansion to the stencil Library Node, with no further changes to the surrounding infrastructure.

\begin{figure}
    \centering
    \begin{minipage}[b]{.32\columnwidth}
        \centering
    \begin{tikzpicture}
        \node[anchor=south west,inner sep=0] at (0, 0) {\includegraphics[height=.3\textheight]{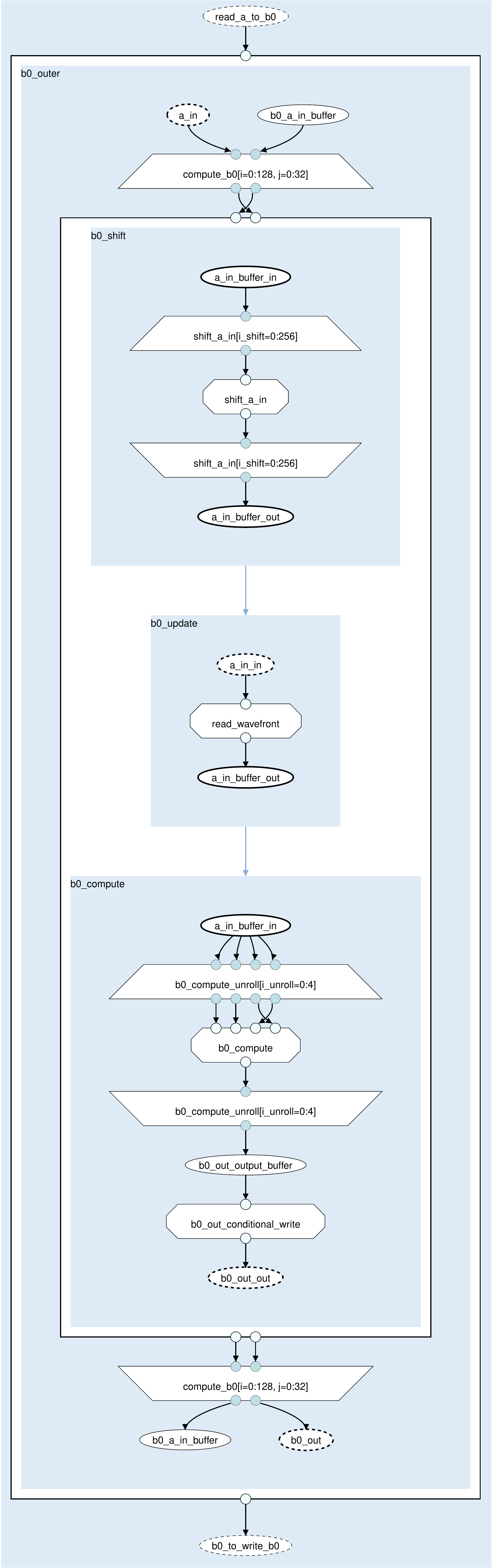}};
        \draw [blue, thick, rounded corners, dashed] (0.5, 4.85) rectangle (1.8, 6.25);
        \draw [red, thick, rounded corners, dashed] (0.8, 3.65) rectangle (1.5, 4.4);
        \draw [black, thick, rounded corners, dashed] (0.5, 1.25) rectangle (1.8, 3.15);
    \end{tikzpicture}
    \end{minipage}\hfill%
    \begin{minipage}[b]{.66\columnwidth}
        \centering
    \begin{tikzpicture}
        \node[anchor=south west,inner sep=0] at (0, 0) {\includegraphics[height=.3\textheight]{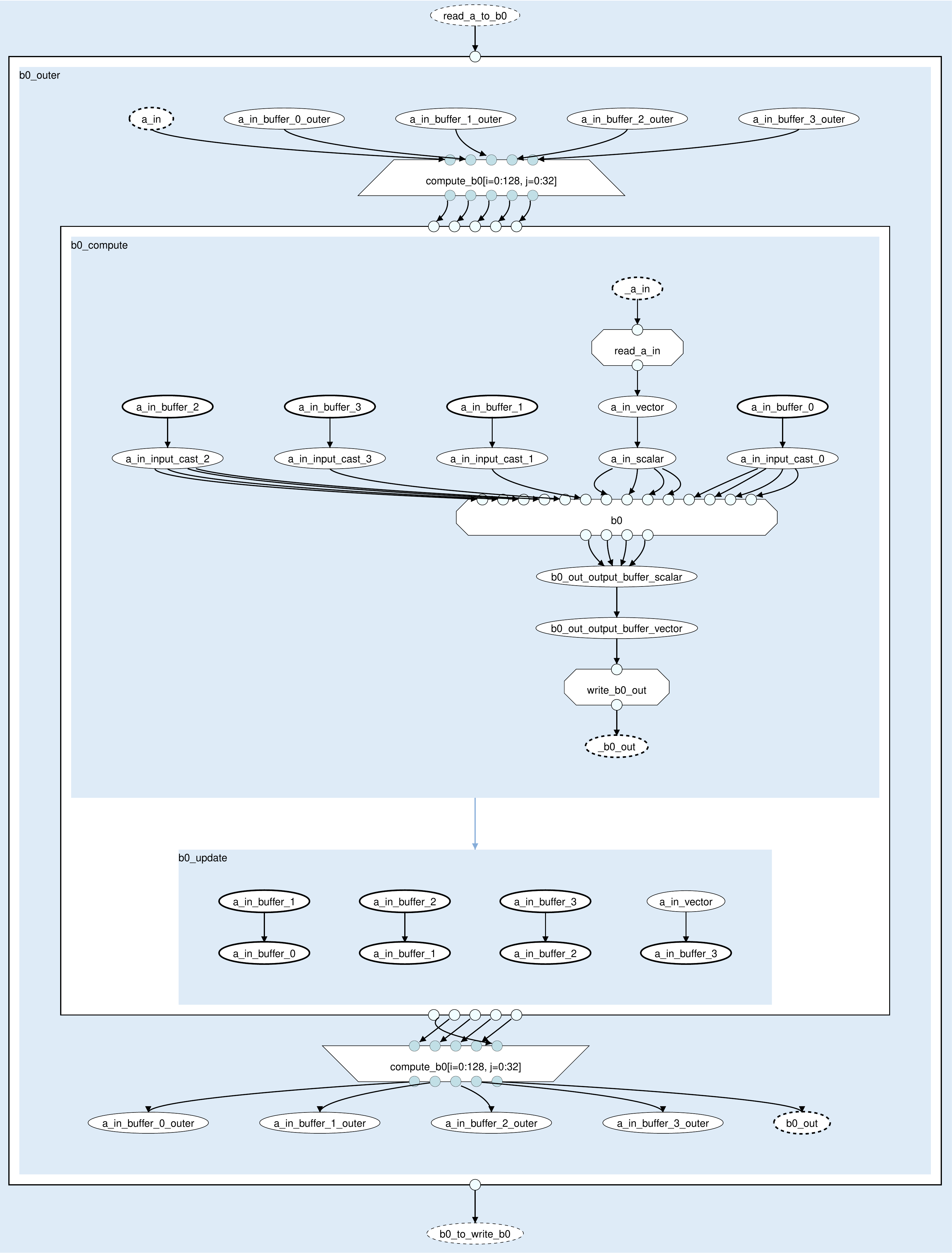}};
    \end{tikzpicture}
    \end{minipage}
    \caption{Intel (left) and Xilinx (right) stencil node expansions.}
    \label{fig:stencilflow_expansions}
\end{figure}
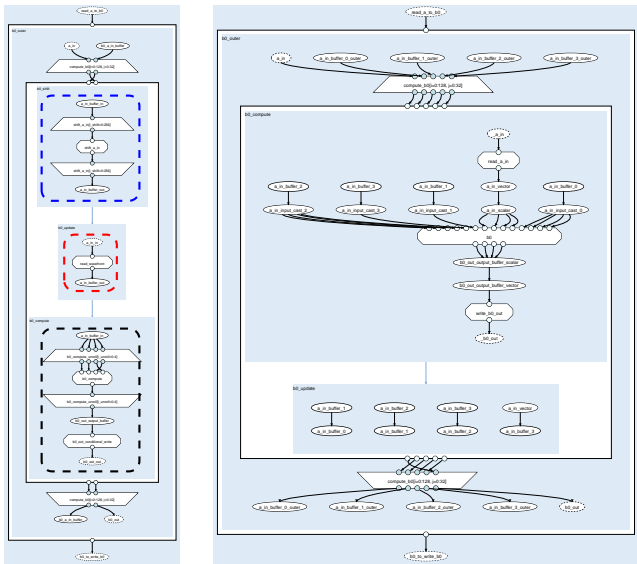

Intel OpenCL's shift register abstraction leaves buffer management for stencil programs up to the compiler. The stencil node expansion subgraph for a 2D 4-point stencil with 4-way vectorization is shown in the \emph{left} graph in Figure~\ref{fig:stencilflow_expansions}. At every iteration, the shift register buffer is ``shifted'' forward by an amount equivalent to the vector length (blue box, top). Then, a new input vector of data from the wavefront is written to the front of the shift register buffer (red box, middle), resulting in the shift register containing all data necessary to perform the stencil computation. Finally, the computation is performed in an unrolled map over the vector width (black box, bottom), loading the $4{\times}4$ appropriate entries from the shift register (four accesses in each iteration of the size-4 unrolled map, although some will overlap in practice), and the result is written to the output stream.

Xilinx does not expose a scalable shift register abstraction. Instead, the buffers between each stencil access must be deduced, instantiated, and accessed explicitly, which is challenging due to vectorization resulting in non-aligned accesses into the vector strides. The \emph{right} graph in Figure~\ref{fig:stencilflow_expansions} shows a Xilinx expansion, achieving the exact same computation as the left graph, but without the aid of shift registers. The 4-point stencil has four access points, which with 4-way vectorization requires $14$ unique offset. Offsets are computed as the distance from the ``earliest access'' relative to the iteration pattern. These offsets are translated into major indices (which buffer is accessed, according to the vector stride) and minor indices (indices into each accessed vector). The major indices become the ``access points'' into the vectorized buffers, resulting in 4 buffers for this example. At each iteration, the buffers are read at a cyclic index along with the value from the wavefront. Because the kernel is vectorized, a full vector must be read from each buffer first, after which the individual scalar elements can be extracted by the kernel (implemented by the $14$ dataflow edges between the buffers and the computation). Finally, each buffer is updated with the value from the \emph{following} access point, and the front access point (i.e., highest flattened index) is updated from the wavefront.

While it required a non-trivial effort to implement a Library Node expansion for Xilinx that implements a stencil buffering pattern, the payoff is significant: the StencilFlow frontend and analysis framework, as well as the remaining SDFG (the memory readers/writers, and the interconnection between stencils) remains unchanged, and we can now directly compile it for Xilinx instead.

\subsection{Evaluation}

We evaluate the Xilinx stencil expansion on the Alveo U250 accelerator board from kernels produced by the StencilFlow framework. While the Alveo U250 is a flagship UltraScale+ device, it suffers relative to Intel chips in not having native floating point units, and from its chiplet-based design, which limits connectivity between parts of the device. Still, the large amount of general purpose logic available allows instantiating large stencil programs.

\begin{figure}
    \centering
    \includegraphics[width=.99\columnwidth]{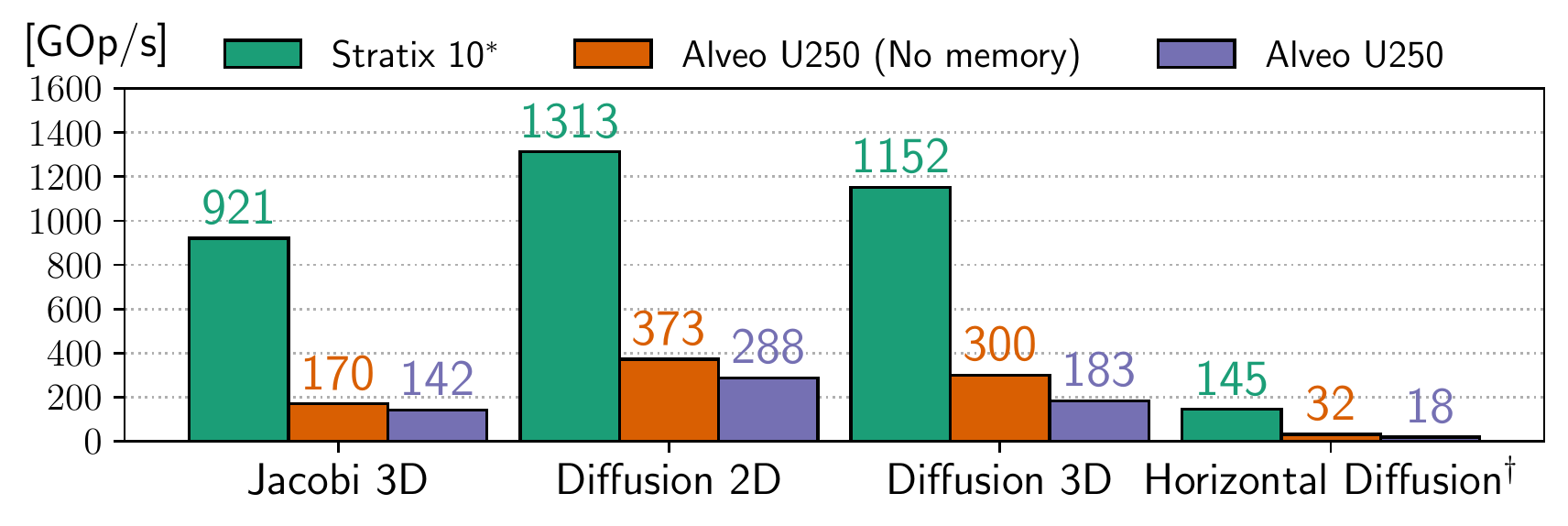}
    \caption{Performance of StencilFlow across Intel and Xilinx platforms.\\$^*$From StencilFlow~\cite{stencilflow} paper. $^\dagger$Preliminary results: not yet validated.}
    \label{fig:stencilflow_performance}
\end{figure}

We benchmark the vectorized Jacobi~3D, diffusion~2D, and diffusion~3D stencil programs on the Xilinx U250 board with 32-bit floating point types, using long and narrow $2^{17}{\times}4096$ and $2^{15}{\times}128{\times}128$ domains for 2D and 3D, respectively, to emulate time tiled stencils. We plot the results along with benchmarks on a Stratix~10 board evaluated by StencilFlow in Figure~\ref{fig:stencilflow_performance}. The plot includes benchmarks both with and without accessing DRAM, as the Alveo board was observed to deliver significantly less than the expected memory bandwidth, despite the burst-friendly access pattern. The U250 yields up to $\SI{373}{\giga\op\per\second}$ ($\SI{300}{\giga\op\per\second}$) without (and with) memory, but falls short of the much larger floating point capabilities of the Stratix~10. The Xilinx stencil performance yields a $3.2{\times}$ improvement over the stencil architecture evaluated in the original work on DaCe~\cite{dace} (which used a KU115 board), and a $2.8{\times}$ over the SODA~\cite{soda} framework (ADM-PCIE-KU3 board). Although both these works use previous generation FPGAs, the domain-specific StencilFlow stack is responsible for a significant benefit.

Finally, we include preliminary results on the U250 for executing \emph{horizontal diffusion}, a large real-life weather simulation stencil program~\cite{stencilflow}, which contains a multitude of heterogeneous stencil computations and buffering patterns with complex dependencies between them. The program is run in a fully pipeline parallel fashion, where every stencil operation is performed in parallel in a streaming fashion, requiring careful buffering on channel connections to avoid deadlocks. Because the buffering is done at the SDFG-level, this can be directly compiled for the Xilinx architecture along with the stencil expansion. The program is applied on a $128{\times}128{\times}80$ grid using 32-bit floating point types. However, both memory and compute utilization is poor, pending further investigation into why the U250's resources are not sufficiently exploited.

\section{Related Work}
\label{sec:related_work}

HeteroCL~\cite{heterocl} proposes a programming model based on a high-level DSL extended from TVM~\cite{tvm}, allowing the user to express programs through a set of computational primitives, some of which are specialized to specific backends: a stencil backend~\cite{soda} and a tensor multiplication backend~\cite{polysa} framework). Optimizations can be done by the programmer by setting parameters of the employed patterns, following the approach of Halide~\cite{halide_fpga}. The SDFG-based method presented here takes a somewhat opposite approach: rather than exposing existing backends through a unified DSL, we unify the full optimization flow and code generation process within a single representation, which can be targeted and customized by any number of internal or external frontends and DSLs. The benefit of HeteroCL is a more ``push-button'' approach: by constraining the input to certain patterns supported by the specialized backends, less intervention is needed by the engineer to achieve fast programs.
In PyLog \cite{pylog}, the programmer writes python code, with the framework being responsible of applying optimization passes and generating C++ annotated code targeting Xilinx devices (through PYNQ \cite{pynq}). 
Compared with both HeteroCL and PyLog, the benefit of SDFGs and the DaCe framework is malleability: every step of the design and optimization process is provided as a tool to the programmer, and every component can be customized, extended, or even hacked with low-level code. Domain-specific constructs and transformations exist within the same space as general constructs and transformations, and can be applied interchangeably. This consequently yields out-of-the-box cross-platform compatibility, as any frontend, DSL, or application based on SDFGs has access to either backend, even if is desirable to tweak details to a specific platform.


Spatial~\cite{spatial} is a language for programming spatial systems through a combination of parallel patterns that constrain the input, and an abstraction of the underlying hardware that these patterns can be efficiently mapped to. The constrained input space is exploited to perform design-space exploration of the hardware mapping, focusing on automated performance tuning of the design. In contrast, the DaCe framework focuses on interactive design based on the data-centric model, which requires more guidance by the performance engineer, but shares its knowledge and toolbox between different systems, and is built to be extensible at every level of the stack.

The multi-level approach taken here is similar to that of MLIR~\cite{mlir}, which introduces multi-level design to compiler IR, where \emph{progressive lowering} allows going from high-level domain-specific constructs~\cite{domain_specific_ir}, through any number of intermediate formats, down to traditional compiler IR such as LLVM. MLIR targets an automated compiler-based approach, while SDFGs are an interactive format that is progressively transformed based on the provided complete view of the program's data movement.

Kenter~et~al.~\cite{portable_opencl} propose a macro-based approach to writing HLS programs that are portable between Xilinx and Intel OpenCL flows. DaCe lifts cross-compatibility into a full code generator, allowing a superset of opportunities to emit highly targeted and structured code for either vendor.

\section{Conclusion}
\label{sec:conclusion}

We proposed leveraging the Stateful DataFlow multiGraphs (SDFGs) representation as a programming model for spatial computing systems. SDFGs express programs by the dataflow and control flow, exposing all data movement to the performance engineer. We showed how the representation is code generated to efficient architectures for either FPGA vendor, and how \emph{Library Nodes} allow embedding abstract domain-specific behavior into the representation, allowing SDFGs to be harnessed using a multi-level design methodology. We showed how programs from linear algebra, machine learning, and stencil domains can be written using expressive high-level frontends, and optimized with both domain-specific transformations, and general purpose transformations reused across domains, then specialized further to exploit platform features. The multi-level methodology promotes the invention of novel transformations and abstractions that further increase productivity and performance not only for the application at hand, but for any future SDFG amenable to the same optimization strategy.

\section*{Acknowledgements}
\begin{wrapfigure}{l}{.3\linewidth}
    \centering
    \includegraphics[width=1\linewidth]{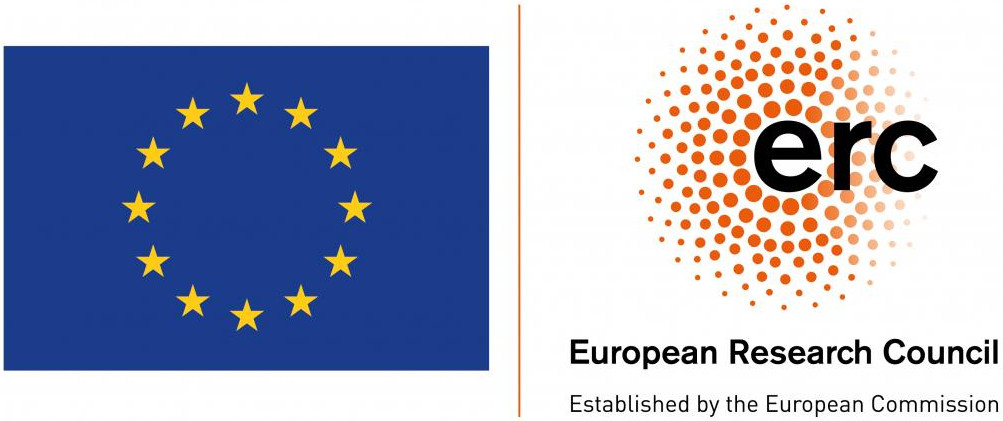}
    \vspace{-10pt}
\end{wrapfigure}
This project received funding from the European Research Council (ERC) grant PSAP, grant agreement No. 101002047, and the European Union’s Horizon Europe programme DEEP-SEA, grant agreement No. 955606.
T.B.N. is supported by the Swiss National Science Foundation (Ambizione Project \#185778).
The project was also sponsored by the Paderborn University, under the DaceML-FPGA
project.
\bibliographystyle{IEEEtran}
\bibliography{refs}




\end{document}